\documentclass[12pt]{article}
\pdfoutput=1
\usepackage{amsmath,amssymb,amsthm,bm,graphicx,float,array,multirow,multicol,rotfloat,caption,subcaption,enumerate,geometry,mathdots,adjustbox,booktabs,parskip,mathtools,tikz,tikz-cd,pdflscape,csquotes,lscape,rotating,empheq,mathrsfs,comment}
\usepackage[para]{threeparttable}
\usepackage[all]{xy}
\usepackage[shortlabels]{enumitem}
\usepackage[normalem]{ulem}
\usepackage[vcentermath]{youngtab}
\usepackage[boxsize=.8em]{ytableau}
\usepackage[numbers,sort&compress]{natbib}
\usepackage[hidelinks]{hyperref}
\usepackage{cleveref}
\usepackage{tensor}

\numberwithin{equation}{section}
\numberwithin{figure}{section}
\allowdisplaybreaks
\makeatletter
\usetikzlibrary{arrows,arrows.meta,shapes,shapes.misc,patterns,decorations.pathmorphing,decorations.pathreplacing,positioning,chains,fit}
\pgfdeclarepatternformonly{coarsedots}{\pgfqpoint{-1pt}{-1pt}}{\pgfqpoint{5pt}{5pt}}{\pgfqpoint{12pt}{12pt}}{
    \pgfpathcircle{\pgfqpoint{.5pt}{.5pt}}{.5pt}
    \pgfusepath{fill}}
\tikzset{gauge/.style={rounded rectangle, draw=black!100, thick, minimum size=5mm},  gaugeD/.style={rounded rectangle, draw=black!100,double,thick,minimum size=5mm},  empty/.style={rounded rectangle, draw=white!100, thick, minimum size=5mm}, flavor/.style={rectangle, draw=black!100, thick, minimum size=5mm},flavorD/.style={rectangle, draw=black!100, double,thick, minimum size=5mm}}
\hypersetup{colorlinks=true}
\hypersetup{linkcolor=black}
\hypersetup{citecolor=black}
\hypersetup{urlcolor=black}
\theoremstyle{plain}
\newtheorem*{thm*}{Theorem}

\makeatother
\usetikzlibrary{matrix}

\def\beq#1\eeq{\begin{align}#1\end{align}}
\def\be#1\ee{\begin{equation}#1\end{equation}}

\newcommand{\Dt}{\tilde{\Delta}}
\newcommand{\Vd}{\dot{V}}
\newcommand{\Vdd}{\ddot{V}}

\def\pd{\partial}

\newcommand{\dd}{\mathrm{d}}

\newcommand{\me}{\mathrm{e}}

\newcommand{\D}{\mathcal{D}}
\newcommand{\vol}{\text{vol}}
\newcommand{\lp}{\ell_p}

\newcommand{\M}{\mathcal{M}}
\begin{document}

\begin{titlepage}
\vspace*{-3cm} 
\begin{flushright}
{\tt CALT-TH-2023-053} \\
{\tt DESY-23-078}\\
\end{flushright}
\begin{center}
\vspace{1.5cm}
{\LARGE\bfseries Holographic duals of Higgsed $\mathcal{D}_p^{\,b}(BCD)$\\}
\vspace{1cm}

{\large Christopher Couzens$^1$, Monica Jinwoo Kang$^{2,\,3}$, Craig Lawrie$^4$, and Yein Lee$^5$}\\
\vspace{.8cm}
\par
\vspace{.2cm}
{$^1$ Mathematical Institute, University of Oxford,\\
Oxford, OX2 6GG, U.K.}\par
{$^2$ Department of Physics and Astronomy, University of Pennsylvania\\
Philadelphia, PA 19104, U.S.A.}\par
{$^3$ Walter Burke Institute for Theoretical Physics, California Institute of Technology\\
Pasadena, CA 91125, U.S.A.}\par
{$^4$ Deutsches Elektronen-Synchrotron DESY,\\ Notkestr.~85, 22607 Hamburg, Germany}\par
{$^5$ Department of Physics and Research Institute of Basic Science,\\
Kyung Hee University, Seoul 02447, Republic of Korea}
\vspace{.3cm}

\scalebox{.85}{\tt christopher.couzens@maths.ox.ac.uk, monica6@sas.upenn.edu,}\\

\scalebox{.85}{\tt craig.lawrie1729@gmail.com, lyi126@khu.ac.kr}\par
\vspace{1.2cm}
\textbf{Abstract}
\end{center}
\noindent
We construct the AdS$_5$ holographic duals to all non-Lagrangian 4d $\mathcal{N}=2$ superconformal field theories of Argyres--Douglas type, namely, $\mathcal{D}_p^{\,b}(G)$, arising from class $\mathcal{S}$ of classical type involving irregular punctures of regular semi-simple type. The 11d supergravity duals contain an internal space of the form of a fibered product of a disc with a squashed and fibered four-sphere and includes orbifold projections which depend on the type of twist lines/outer-automorphism twists in the class $\mathcal{S}$ theory. We verify the holographic duality by determining and matching the anomalies (including the central charges $a$ and $c$ and the flavor central charges) at leading and subleading orders. The Higgs branch of the conformal field theory is described via Higgsing by a nilpotent orbit of a classical Lie algebra; we find the exact closed form formulae for the central charges for every Higgsing. We prove that in the supergravity duals, constraints on the type of partitions associated to allowable Higgsings are enforced by the consistency condition known as the t-rule.

\vspace{1cm}
\vfill 
\end{titlepage}

\tableofcontents
\newpage

\section{Introduction}

Non-perturbative dynamics or strong-coupling phenomena are one of the difficult arenas of quantum field theories where traditional perturbative tools fail to be effective.\footnote{There are other techniques such as supersymmetric localization, integrability, and the conformal bootstrap, but they are sometimes not powerful enough or simply not applicable.} In particular, Argyres--Douglas theories are well-known four-dimensional examples whose behavior is inherently strongly-coupled. 
Holography, being a strong-weak duality, is a powerful technique that may be used to extract non-perturbative results. A renowned example is where the strong-coupling behaviour of the anomalous dimension of unprotected scalar operators in 4d $\mathcal{N}=4$ super Yang--Mills is determined from computing masses of particles in classical supergravity \cite{Gubser:1998bc}. As such, it is a useful and important program to identify holographic pairs, on the gravity side a $(d+1)$-dimensional Anti-de-Sitter solution and on the field theory side the dual $d$-dimensional superconformal field theory. 

To provide evidence for a putative holographic pair, it is imperative to have some properties sufficiently under control that they can be determined from both the gravitational and field theoretic perspectives. While the strong-weak nature of the holographic duality guarantees that we have computational control in the gravity description, the strongly-coupled nature of the field theory precludes any straightforward, bottom-up analysis. Instead, pioneered from the realization of the non-Lagrangian 6d $(2,0)$ SCFTs in Type IIB string theory \cite{Witten:1995zh}, a powerful technique has emerged whereby strongly-coupled QFTs are realized top-down, via either higher-dimensional field theories or string theory.\footnote{For a recent review of the potency of this top-down perspective for constructing QFTs, see \cite{Argyres:2022mnu}.} One such approach to strongly-coupled 4d $\mathcal{N}=2$ SCFTs is known as class $\mathcal{S}$ \cite{Gaiotto:2009we,Gaiotto:2009hg}; this involves starting from the 6d $(2,0)$ SCFTs and further compactifying on a punctured Riemann surface to four dimensions. Many of the intricate non-perturbative features of the resulting 4d SCFTs, such as S-dualities, are geometrized; they are encoded in the geometric properties of the punctured Riemann surface. The richness of class $\mathcal{S}$ is related to the wide variety of possible punctures; punctures can be regular, irregular, twisted, untwisted, etc, and the different types of punctures lead to qualitatively different physical features.

A particularly interesting set of strongly-coupled 4d $\mathcal{N}=2$ SCFTs are the Argyres--Douglas theories. These are a class of theories that generically have Coulomb branch operators with non-integer scaling dimensions. Argyres--Douglas SCFTs were originally obtained by studying special points on the moduli space of 4d $\mathcal{N}=2$ gauge theories where dyonic states that are mutually non-local become simultaneously massless \cite{Argyres:1995jj,Argyres:1995xn,Eguchi:1996vu}. In this paper, we will generally use the term ``Argyres--Douglas theories'' to refer to class $\mathcal{S}$ theories with an irregular puncture, as these theories typically have Coulomb branch operators with non-integer scaling dimensions. Due to the fractional nature of the scaling dimensions, these theories lack a 4d $\mathcal{N}=2$ Lagrangian description,\footnote{However, realizations as an infrared fixed point of a 4d $\mathcal{N}=1$ Lagrangian QFT may exist \cite{Maruyoshi:2016tqk,Maruyoshi:2016aim,Agarwal:2016pjo,Agarwal:2017roi,Benvenuti:2017bpg}.} and they are inherently strongly-coupled.

Due to the strongly-coupled nature of Argyres--Douglas theories, and of class $\mathcal{S}$ theories in general, it is important to have a construction for these SCFTs with which we can extract physical properties. These constructions may be from a higher-dimensional field theory, such as the class $\mathcal{S}$ approach that we have just described, from string theory, for example via geometric engineering in Type IIB, or from holography. It turns out that different 
top-down constructions for the same QFT can make manifest different features: physics that is complicated to extract from one description may be obvious from another.\footnote{This plurality of origins has recently been exploited in \cite{Baume:2021qho,Distler:2022yse,Distler:2022kjb}, for example.} In this paper, we propose that certain $\operatorname{AdS}_5 \times X_6$ solutions in 11d supergravity constitute the holographic duals of the Argyres--Douglas theories. We demonstrate the consistency of these putative holographic duals by matching various physical properties, such as the central charges, which are determined independently from the 11d supergravity solution and field theoretically from the class $\mathcal{S}$ perspective.

The Argyres--Douglas theories are not an unstructured collection of theories, but they form themselves into families related by Higgs branch renormalization group flows. Each family possess a parent theory; in the cases we study they are commonly known as $\mathcal{D}_p^{\,b}(G)$. Here $G$ is a simple Lie algebra which is a flavor symmetry of the parent theory, and all other theories in the family can be obtained as the infrared fixed point of a Higgs branch RG flow triggered via giving a nilpotent vacuum expectation value to the moment map of that flavor symmetry. Therefore, each Argyres--Douglas theory is associated to a parent theory, $\mathcal{D}_p^{\,b}(G)$, and a nilpotent orbit $O$ of $G$ that describes the Higgsing. The choice of nilpotent orbit, which for $G$ a classical Lie algebra corresponds to some integer partition/Young tableau, must be encoded in the $\operatorname{AdS}_5 \times X_6$ supergravity solution. The nilpotent orbits of $G$ possess a partial ordering, which coincides with the hierarchy of RG flows between the associated 4d $\mathcal{N}=2$ SCFTs. In fact, as for the analogous 6d $(1,0)$ setup described in \cite{DeLuca:2018zbi}, we expect the existence of an 11d supergravity solution for each family, where the supersymmetric $\operatorname{AdS}_5$ vacua correspond to each SCFT in the family, and where domain wall solutions connecting two such vacua exist if and only if the two SCFTs are related via a Higgs branch RG flow. 

While holographic duals of class $\mathcal{S}$ theories with only regular punctures have been studied previously \cite{Gaiotto:2009gz}, the holographic duals of Argyres--Douglas theories have remained elusive until recently. In \cite{Bah:2021hei,Bah:2021mzw} it was proposed that the holographic dual of the $\mathcal{D}_p^{\,b}(A)$ theory with rectangular\footnote{A rectangular puncture is one associated to a rectangular Young tableau.} regular puncture was given by the uplift of an AdS$_5\times \mathbb{D}_2$ solution of 7d maximal gauged supergravity with $\mathbb{D}_2$ a two-dimensional disc.\footnote{A similar research line which evolved in parallel with the construction of disc solutions were spindle solutions, \cite{Ferrero:2020laf,Ferrero:2020twa,Hosseini:2021fge,Boido:2021szx,Faedo:2021kur,Ferrero:2021wvk,Cassani:2021dwa,Ferrero:2021ovq,Couzens:2021rlk,Faedo:2021nub,Ferrero:2021etw,Couzens:2021cpk,Giri:2021xta,Couzens:2022agr,Cheung:2022wpg,Suh:2022olh,Arav:2022lzo,Couzens:2022yiv,Couzens:2022aki,Boido:2022mbe,Suh:2022pkg,Suh:2023xse,Amariti:2023mpg,Kim:2023ncn,Hristov:2023rel,Amariti:2023gcx}, and their four-dimensional generalisations \cite{Cheung:2022ilc,Couzens:2022lvg,Faedo:2022rqx}. A spindle is the weighted projective space $\mathbb{WCP}^1_{n_\pm}$, that is it is topologically a sphere with conical singularities at both poles. Discs can be understood as different global completions of the same local solutions which furnish spindles. No field theory duals of spindle solutions are known to date. Another related class of solutions are the defect solutions of \cite{Gutperle:2022pgw,Gutperle:2023yrd}.} The disc has a conical singularity at its center and a boundary at which the metric is singular. Once uplifted to 11d the conical singularity gives rise to a $\mathbb{R}^4/\mathbb{Z}_n$ singularity while the singular boundary becomes the locus of a smeared stack of M5-branes. In the dual field theory, the  $\mathbb{R}^4/\mathbb{Z}_n$ singularity is dual to a regular puncture while the smeared M5-brane is the holographic dual of an irregular puncture. As such, the disc may be better understood as a sphere with one regular puncture and one irregular puncture at the opposite pole exactly as in the class $\mathcal{S}$ construction.\footnote{Disc solutions have also been constructed in different theories, corresponding to the compactification of different SCFTs on a punctured sphere, see \cite{Couzens:2021tnv,Suh:2021ifj,Suh:2021aik,Suh:2021hef,Couzens:2021rlk,Karndumri:2022wpu,Bomans:2023ouw}.}

Later in \cite{Couzens:2022yjl, Bah:2022yjf} the holographic duals of the $\mathcal{D}_p^{\,b}(A)$ theories with arbitrarily Higgsed flavor symmetry were constructed. The key observation in \cite{Couzens:2022yjl, Bah:2022yjf} was that the solutions could be embedded into the refined classification of $\mathcal{N}=2$ AdS$_5$ solutions of \cite{Gaiotto:2009gz}. The solutions are determined by a potential satisfying the 3d cylindrical Laplace equation, and using the linearity of the PDE, the most general solutions with arbitrary Higgsing were constructed. One of the novel features of these solutions is that they evade the quantization conditions imposed in \cite{Gaiotto:2009gz} which lead to Lagrangian quiver theory duals, generalising to the inclusion of fractional M5-branes and thus non-Lagrangian duals.

The goal of this work is to identify the holographic duals of the (arbitrarily Higgsed) $\mathcal{D}_p^{\,b}(BCD)$ theories. As we will see, this requires the introduction of $\mathbb{Z}_2$ orbifolds, or from a Type IIA perspective O4-orientifold planes. The presence of the orbifolds imposes additional consistency conditions which are dual to the conditions for a BCD-partition to be well-defined. Moreover the presence of the orbifolds leads to a non-trivial Stiefel--Whitney class for a subset of the orbifolds which requires both a modification of the Dirac quantisation condition and anomaly inflow. A similar investigation for orientifolds of 6d $\mathcal{N}=(1,0)$ SCFTs in massive Type IIA was performed in \cite{Apruzzi:2017nck}. There they managed to match the observables to leading order and it would be interesting to include the shifts we introduce here there too in order to match to subleading order.

The structure of this paper is as follows. In Section \ref{sec:nilphiggsing}, we analyze the anomalies for 4d $\mathcal{N}=2$ SCFTs obtained via nilpotent Higgsing of a non-Abelian flavor factor of some ultraviolet SCFT. Then, in Section \ref{sec:Fieldtheory}, we review the class $\mathcal{S}$ construction and introduce the $\mathcal{D}_p^{\,b}(ABCD)$ SCFTs associated to two-punctured spheres with one regular maximal puncture and one irregular puncture of regular semi-simple type. In Section \ref{sec:holoDpG}, we construct the supergravity solutions in the eleven-dimensional theory, which is dual to $\mathcal{D}_p^{\,b}(BCD)$ theory and match the anomaly coefficients for $\mathcal{D}_p^{\,b}(ABCD)$ theory obtained via nilpotent Higgsing using the anomaly inflow method. To exemplify our results, in particular the matching of the anomalies between the SCFTs and the holographic duals, we focus in Section \ref{sec:rect} on $\mathcal{D}_p^{\,b}(ABCD)$ theories where the regular maximal puncture is Higgsed by a nilpotent orbit associated to a rectangular partition. Finally, we discuss our results and provide some avenues for future research in Section \ref{sec:disc}.

\section{4d \texorpdfstring{\boldmath{$\mathcal{N}=2$}}{N=2} SCFT Higgs branch via nilpotent Higgsing}\label{sec:nilphiggsing}

In this section, we study how a 4d $\mathcal{N}=2$ SCFT with a flavor symmetry factor $\mathfrak{g}$ behaves under nilpotent Higgsing of $\mathfrak{g}$. Consider a generic 4d $\mathcal{N}=2$ SCFT with a flavor symmetry algebra $\mathfrak{t}$ that contains a simple Lie algebra factor $\mathfrak{g}$ such that
\begin{align}
    \mathfrak{t}=\mathfrak{g}\oplus\mathfrak{h} \,,
\end{align}
where $\mathfrak{h}$ is either trivial, $\mathfrak{u}(1)$, a simple Lie algebra, or a sum of them. Due to the existence of the flavor symmetry, there exists a moment map operator (i.e., a scalar primary) $\mu$ inside of a $\widehat{\mathcal{B}}_1$ supermultiplet,\footnote{We use the notation of \cite{Dolan:2002zh} for unitary irreducible representations of the 4d $\mathcal{N}=2$ superconformal algebra.} which transforms in the adjoint representation of $\mathfrak{g}$. We can consider Higgsing by giving a vacuum expectation value to the moment map as
\begin{equation}
    \langle \mu \rangle = J \,,
\end{equation}
where we have chosen a vev proportional to a nilpotent element $J$ of $\mathfrak{g}$.\footnote{It is not necessarily true that it is always possible to turn on a vev in this way, for example, there may exist chiral ring conditions that forbid certain nilpotent vevs.} The resulting infrared theory depends only on the conjugacy class of the nilpotent element, i.e., on a nilpotent orbit $O$ fixed by $J$. 

From this setup we now demonstrate explicitly how the 't Hooft anomalies (and thus central charges or conformal anomalies) of the Higgsed theory can be expressed in terms of the anomalies of unHiggsed theory and the properties of the nilpotent orbit $O$. 
While in this paper, we will apply these results to the $\mathcal{D}_p^{\,b}(G)$ SCFTs, the results of this section hold for many 4d $\mathcal{N}=2$ SCFTs with a flavor symmetry $\mathfrak{g}$. To be more specific, the results in this section hold for any 4d $\mathcal{N}=2$ SCFT with flavor symmetry $\mathfrak{g}$ where the nilpotent Higgsing of $\mathfrak{g}$ only decouples the Nambu--Goldstone modes coming from the flavor current multiplet. In particular, if $\mathfrak{g}$ arises as the flavor symmetry of a maximal puncture in class $\mathcal{S}$, then the results hold. Some counterexamples where additional fields decouple can be seen in \cite{Distler:2022jvk, DEKL}. In this section, we follow the algorithm to determine the Nambu--Goldstone modes presented in \cite{Tachikawa:2015bga}.

For a simple non-Abelian Lie algebra $\mathfrak{g}$, it is a standard result that nilpotent orbits in $\mathfrak{g}$ are in one-to-one correspondence with conjugacy classes of homomorphisms  $\rho: \mathfrak{su}(2)_X \rightarrow \mathfrak{g}$.\footnote{For a comprehensive review of nilpotent orbits, see \cite{MR1251060}.} Let $\mathfrak{f}$ denote centralizer of $\rho$, that is, the Lie subalgebra in $\mathfrak{g}$ which commutes with the image of $\rho$. Then each nilpotent orbit can be associated to a distinct decomposition of $\mathfrak{g}$:\footnote{When the nilpotent orbit is the maximal nilpotent orbit, then $\mathfrak{su}(2)_X$ is embedded trivially in $\mathfrak{g}$, and the centralizer is then all of $\mathfrak{g}$. We are abusing notation throughout this paper by using $\mathfrak{su}(2)_X$ to denote both the domain and the image of $\rho$.}
\begin{equation}\label{eqn:su2Xembed}
    \mathfrak{g} \rightarrow \mathfrak{su}(2)_X \oplus \mathfrak{f} \,.
\end{equation}
We can write $\mathfrak{f}$ as
\begin{equation}\label{eqn:fj}
    \mathfrak{f} = \bigoplus_i \mathfrak{j}_i \,,
\end{equation}
where $\mathfrak{j}_i$ is either trivial, $\mathfrak{u}(1)$, or a simple Lie algebra.

Nilpotent Higgsing of a flavor symmetry $\mathfrak{g}$ corresponds to a Higgs branch renormalization group flow. As such, the $\mathfrak{su}(2)_R$ R-symmetry of the ultraviolet theory, under which the moment map transforms in the $\bm{3}$, is broken along the flow. At the infrared fixed point a new $\mathfrak{su}(2)_R'$ R-symmetry emerges; this is simply the diagonal combination of the original $\mathfrak{su}(2)_R$ and the $\mathfrak{su}(2)_X$. Furthermore, the UV flavor symmetry $\mathfrak{g}$ is broken to $\mathfrak{f}$ in the IR. To determine the infrared 't Hooft anomaly coefficients, we want to know the Nambu--Goldstone modes that arise from the ultraviolet moment map operator. This procedure is explained in \cite{Tachikawa:2015bga}. In particular, we should consider the branching rule of the adjoint representation of $\mathfrak{g}$ under the decomposition in equation \eqref{eqn:su2Xembed}:
\begin{equation}\label{eqn:adjdecomp}
    \textbf{adj} \quad \rightarrow \quad \bigoplus_\ell V_\ell \otimes R_\ell \,,
\end{equation}
where $V_\ell$ is the irreducible $\mathfrak{su}(2)$ representation of dimension $d_\ell$ and $R_\ell$ is some irreducible representation of $\mathfrak{f}$. Using the decomposition in equation \eqref{eqn:fj}, the representation $R_\ell$ can be decomposed as
\begin{equation}
    R_\ell = \bigotimes_i R_\ell^{(i)} \,.
\end{equation}

Under the assumption that the infrared theory is a good SCFT, the Nambu--Goldstone modes include Weyl fermions transforming in the $(d_\ell - 1)$-dimensional representation of $\mathfrak{su}(2)_R'$, with R-charge $-1$ under the $\mathfrak{u}(1)_r$, and in the representation $R_\ell$ of $\mathfrak{f}$, for each entry in the sum in equation \eqref{eqn:adjdecomp}. There are also bosonic Nambu--Goldstone bosons which combine with the Weyl fermions to form Nambu--Goldstone hypermultiplets; however, as we are primarily interested in the anomalies, we do not go into the details of these bosonic modes. 

We now determine the contributions to the anomalies from these Nambu--Goldstone modes. As usual, 
a chiral fermion transforming in a representation $\rho$ of the global symmetry algebra contributes to the six-form anomaly polynomial as
\begin{equation}
    \widehat{A}(T) \operatorname{tr}_\rho e^{iF} \, \big|_\text{6-form} \,\,,
\end{equation}
where $T$ is the tangent bundle to 4d spacetime and $F$ is the curvature of the global symmetry. 

To perform the expansion of the characteristic classes, we need to recall a few facts on characteristic classes and how to convert traces between different representations. The one-instanton normalized trace for a simple Lie algebra $\mathfrak{g}$ can be defined as 
\begin{equation}
    \operatorname{Tr} F^2 = \frac{1}{h^\vee_\mathfrak{g}} \operatorname{tr}_\textbf{adj} F^2 \,.
\end{equation}
To convert the one-instanton normalized trace to the trace in an arbitrary representation $\rho$ of $\mathfrak{g}$, we define the coefficient $A_\rho$ such that
\begin{equation}
    \operatorname{tr}_\rho F^2 = A_{\rho} \operatorname{Tr} F^2 \,.
\end{equation}
The one-instanton normalized trace is related to the second Chern class of the symmetry using
\begin{equation}
    c_2(F) = \frac{1}{4} \operatorname{Tr} F^2 \,.
\end{equation}
For $\mathfrak{g} = \mathfrak{su}(2)$, the trace in the $d$-dimensional irreducible representation $\bm{d}$ can be written in terms of the one-instanton normalized trace as follows:
\begin{equation}
    \operatorname{tr}_{\bm{d}}F^2 = A_{\bm{d}}\operatorname{Tr} F^2 = \frac{d(d^2 - 1)}{12} \operatorname{Tr} F^2 \,.
\end{equation}
We also remind the reader that the expansion of the A-roof genus is
\begin{equation}
    \widehat{A}(T) = 1 - \frac{1}{24}p_1(T) + \cdots \,,
\end{equation}
where the $\cdots$ indicates terms of higher form-degree.

It is now straightforward to expand the characteristic classes and find the anomaly polynomial contribution from the Nambu--Goldstone modes we have specified. We find that the result can be written as
\begin{equation}
    I_6^\text{NG} = H\left( \frac{1}{12}c_1(r)p_1(T) - \frac{1}{3}c_1(r)^3 \right) + H_v c_1(r)c_2(R') + \sum_i H_i c_1(r) c_2(F_i) \,,
\end{equation}
where $F_i$ is the field strength for each $\mathfrak{j}_i$, $c_1(r)$ is the first Chern class of the $\mathfrak{u}(1)_r$ R-symmetry bundle, and $c_2(R')$ is the second Chern class of the bundle associated to the $\mathfrak{su}(2)_{R'}$ symmetry. The coefficients are as follows:
\begin{equation}\label{eqn:COOLHs}
    \begin{gathered}
        H = \frac{1}{2} \sum_{\ell} (d_\ell - 1)  \operatorname{dim} R_\ell \,, \qquad
        H_v = \frac{1}{6} \sum_{\ell} d_\ell (d_\ell - 1) (d_\ell - 2) \operatorname{dim} R_\ell \,, \\
        H_i = 2 \sum_{\ell} (d_\ell - 1) A_{R_\ell^{(i)}}\frac{\operatorname{dim} R_\ell}{\operatorname{dim} R_\ell^{(i)}} \,.
    \end{gathered}
\end{equation}

Now that we have elucidated the spectrum of Nambu--Goldstone modes and their anomaly polynomial, we explain how they affect the anomaly coefficients of the infrared SCFT after Higgsing. The 't Hooft anomalies/central charges are conveniently encoded in the formal six-form anomaly polynomial, $I_6$, written in terms of the characteristic classes of the symmetry bundles of the theory. The anomaly polynomial of the ultraviolet theory with flavor factor $\mathfrak{g}$ takes the form\footnote{We ignore anomalies for Abelian flavor symmetries throughout this paper.}
\begin{equation}\label{eqn:UVAP}
  \begin{aligned}
    I_6 &= 24(c^\text{UV} - a^\text{UV})\left(\frac{1}{12}c_1(r)p_1(T) - \frac{1}{3}c_1(r)^3 \right) - 4(2a^\text{UV} - c^\text{UV}) c_1(r)c_2(R) \\&\qquad\qquad\qquad\qquad\qquad\qquad\qquad + k_G^\text{UV} c_1(r) c_2(F) + \sum_a k_a^\text{UV} c_1(r) c_2(H_a) \,,
  \end{aligned}
\end{equation}
where $c_1(r)$ and $c_2(R)$ are, respectively, the Chern classes of the $\mathfrak{u}(1)$ and $\mathfrak{su}(2)_R$ R-symmetries; $p_1(T)$ is the first Pontryagin class of the tangent bundle to spacetime; $c_2(G)$ is the second Chern class of the flavor symmetry $G$; and $c_2(H_a)$ is the Chern class of any additional simple non-Abelian flavor symmetry factors, indexed by $a$. After Higgsing via the nilpotent orbit $O$ of $\mathfrak{g}$, we obtain the anomaly polynomial for the infrared theory:
\begin{equation}\label{eq:AMIR}
  \begin{aligned}
    I_6 &= 24(c^\text{IR} - a^\text{IR})\left( \frac{1}{12}c_1(r)p_1(T) - \frac{1}{3}c_1(r)^3\right) - 4(2a^\text{IR} - c^\text{IR}) c_1(r)c_2(R') \\&\qquad\qquad\qquad\qquad\qquad\qquad\qquad + \sum_i k_i^\text{IR} c_1(r) c_2(F_i) + \sum_a k_a^\text{IR} c_1(r) c_2(H_a) \,,
  \end{aligned}
\end{equation}
where $i$ runs over the non-Abelian factors $\mathfrak{j}_i$ in equation \eqref{eqn:fj}, and $c_2(F_i)$ is the second Chern class of the bundle associated to $\mathfrak{j}_i$. To stress that the infrared R-symmetry is different from the ultraviolet R-symmetry, we have written $c_2(R')$ instead of the $c_2(R)$ that appears in equation \eqref{eqn:UVAP}.

Using the fact that the infrared $\mathfrak{su}(2)_R'$ R-symmetry is the diagonal of the UV R-symmetry and the $\mathfrak{su}(2)_X$, together with the anomaly contributions of the Nambu--Goldstone modes, we can determine how the IR 't Hooft anomaly coefficients are related to the UV 't Hooft anomaly coefficients. The final result is\footnote{We emphasize once again that these formulae only hold when the only modes that decouple under the nilpotent Higgsing are the Nambu--Goldstone modes from the moment map that we described.}
\begin{equation}\label{eqn:tachiGB}
    \begin{aligned}
        24(c^\text{IR} - a^\text{IR}) &= 24(c^\text{UV}-a^\text{UV}) - H \,, \\
        4(2a^\text{IR} - c^\text{IR}) &= 4(2a^\text{UV} - c^\text{UV}) - I_X k_G^\text{UV} + H_v \,, \\
        k_i^\text{IR} &= I_i k_G^\text{UV} - H_i \,, \\
        k_a^\text{IR} &= k_a^\text{UV} \,.
    \end{aligned}
\end{equation}

It was stated in \cite{Tachikawa:2015bga} that this analysis in terms of the Nambu--Goldstone modes reproduces the known results \cite{Chacaltana:2012zy} for the 't Hooft anomalies for the infrared theory obtained via partial closure of the maximal puncture in class $\mathcal{S}$. As the $H$, $H_i$, $H_v$, $I_i$, and $I_X$ are straightforward to determine from the perspective of the integer partition underlying the classical nilpotent orbits, we use equation \eqref{eqn:tachiGB} extensively. In the remainder of this section, we work out these quantities explicitly in terms of the integer partitions associated to the nilpotent orbits of the classical simple Lie algebras.

\subsection{\texorpdfstring{$G = SU(N)$}{G = SU(N)}}\label{sec:SU}

For $SU(N)$, nilpotent orbits are in one-to-one correspondence with integer partitions of $N$. We can write such a partition as
\begin{equation}\label{eqn:sugenericpart}
    [1^{m_1}, \cdots, N^{m_N}] \,,
\end{equation}
such that
\begin{equation}
    \sum_i i m_i = N \,.
\end{equation}
Each nilpotent orbit, $O$, is associated to an embedding of $\mathfrak{su}(2)$ in $\mathfrak{su}(N)$. We call the centralizer of this embedding $\mathfrak{f}$, and therefore each nilpotent is associated to a decomposition
\begin{equation}
    \mathfrak{su}(N) \rightarrow \mathfrak{su}(2)_X \oplus \mathfrak{f} \,.
\end{equation}
The centralizer of the $\mathfrak{su}(2)_X$ embedding is given in terms of the nilpotent orbit, and thus of the partition in equation \eqref{eqn:sugenericpart}:\footnote{Of course, $\mathfrak{su}(0)$ and $\mathfrak{su}(1)$ are considered trivial, and only included here for notational convenience. In addition, we choose to only write the non-Abelian simple factors in the centralizer in equation \eqref{eqn:fsu}.}
\begin{equation}\label{eqn:fsu}
    \mathfrak{f} = \bigoplus_{i=1}^N \mathfrak{su}(m_i) \,.
\end{equation}
The Dynkin embedding indices of each of the simple factors in this decomposition of $\mathfrak{su}(N)$ is as follows:
\begin{equation}\label{eqn:SUIX}
    I_X = \frac{1}{6} \sum_{i=1}^N m_i i (i^2 - 1) \,, \qquad I_i = i \,,
\end{equation}
where $I_X$ is the index for $\mathfrak{su}(2)_X$ and $I_i$ for $\mathfrak{su}(m_i)$. The Dynkin index of the embedding can be determined from the branching rule of the fundamental representation of $\mathfrak{su}(N)$, which is directly read off from the partition in equation \eqref{eqn:sugenericpart}:
\begin{equation}\label{eqn:sufunddecomp}
    \bm{N} \quad \longrightarrow \quad \bigoplus_{i=1}^{N} V_i \otimes R_i \,.
\end{equation}
Here, $V_i$ is the $i$-dimensional representation of $\mathfrak{su}(2)_X$ and $R_i$ is the fundamental representation of $\mathfrak{su}(m_i)$.\footnote{More precisely, $R_i$ is a representation of $\mathfrak{f}$ obtained by tensoring the fundamental representation of $\mathfrak{su}(m_i)$ with the trivial representation of each $\mathfrak{su}(m_{j\neq i})$.} To determine the Nambu--Goldstone modes, we need to know about the decomposition of the adjoint representation of $\mathfrak{su}(N)$, which is straightforward to understand from the decomposition of the fundamental in equation \eqref{eqn:sufunddecomp} using:
\begin{equation}
    \bm{N} \otimes \overline{\bm{N}} = \textbf{adj} \oplus \bm{1} \,.
\end{equation}
Recalling that the tensor product of $\mathfrak{su}(2)$ irreps decomposes into a direct sum of irreps as follows
\begin{equation}
    V_i \otimes V_j = \bigoplus_{k=1}^{\operatorname{min}(i,j)} V_{i + j + 1 - 2k} \,,
\end{equation}
we can directly read off the branching of the adjoint representation:
\begin{equation}
    \bm{N} \otimes \overline{\bm{N}} \,\,\rightarrow\,\, \bigoplus_{i,j=1}^N \bigoplus_{k=1}^{\operatorname{min}(i,j)} (V_{i + j + 1 - 2k}, R_i \otimes \overline{R}_j) \,.
\end{equation}
Thus, we can use equation \eqref{eqn:COOLHs} to determine the coefficients appearing in the Nambu--Goldstone mode anomaly polynomial. In the end, we find that the general form, depending on the choice of partition, is
\begin{equation}\label{eqn:SUNGHs}
    \begin{aligned}
        H &= \frac{1}{2}\sum_{i,j=1}^N m_i m_j (ij - \operatorname{min}(i,j)) \,, \\
        H_v &= \frac{1}{6}\sum_{i,j=1}^N m_i m_j (ij - \operatorname{min}(i,j))(i(i-2) + j(j-2) + 2\operatorname{min}(i,j) - 1) \,, \\
        H_i &= 2 \sum_{j=1}^N m_j (ij - \operatorname{min}(i,j)) \,.
    \end{aligned}
\end{equation}
The expression for $H_i$ of course only makes sense when $m_i > 1$; it is the contribution to the anomaly polynomial of the Nambu--Goldstone modes that are charged non-trivially under the $\mathfrak{su}(m_i)$ infrared flavor symmetry factor. To determine $H_i$, we had to use the following trace coefficients for $\mathfrak{su}(m_i)$, 
\begin{equation}
    A_{\textbf{fund}} = \frac{1}{2} \,, \qquad A_{\overline{\textbf{fund}}} = \frac{1}{2} \,, \qquad A_{\textbf{adj}} = h^\vee_{\mathfrak{su}(m_i)} \,.
\end{equation}

\subsection{\texorpdfstring{$G = SO(N)$}{G  = SO(N)}}\label{sec:SO}

For $SO(N)$, each nilpotent orbit is associated to an underlying BD-partition, however, unlike for the other classical Lie algebras, there can be two nilpotent orbits associated to the same BD-partition. This occurs only when $N$ is an integer multiple of four, and the BD-partition contains only even integers. The conventional invariants, such as the 't Hooft coefficients are the same for nilpotent orbits with the same underlying partition, and thus we can ignore this subtlety in this section, and just work at the level of the partition.\footnote{To see how to distinguish theories that differ by Higgsing via nilpotent orbits with the same underlying partition, see \cite{Distler:2022nsn,Distler:2022yse,Distler:2022kjb}.} A BD-partition is an integer partition of $N$ which can be written as 
\begin{equation}\label{eqn:BDpart}
    [1^{m_1}, 2^{m_2}, \cdots, N^{m_N}] \quad \text{ such that } \quad \sum_{i=1}^N i m_i = N \,,
\end{equation}
and which furthermore satisfies the condition that 
\begin{equation}
    i \text{ even } \quad \Longrightarrow \quad  m_i \text{ even } \,.
\end{equation}
Each such partition is associated to an embedding of $\mathfrak{su}(2)$ inside $\mathfrak{so}(N)$, and thus to a decomposition 
\begin{equation}\label{eqn:sodecomp}
        \mathfrak{so}(N) \rightarrow \mathfrak{su}(2)_X \oplus \mathfrak{f} \,.
\end{equation}
Here, $\mathfrak{f}$ is the centralizer of the $\mathfrak{su}(2)$ under the embedding associated to the BD-partition. In this case, $\mathfrak{f}$ is a sum of $\mathfrak{so}$ and $\mathfrak{usp}$ factors, as follows:
\begin{equation}
    \mathfrak{f} = \bigoplus_{i=1}^N \mathfrak{j}_i \qquad \text{ where } \qquad \mathfrak{j}_i = \begin{cases}
        \mathfrak{usp}(m_i) \quad &\text{ if } i \text{ even } \\
        \mathfrak{so}(m_i) \quad &\text{ if } i \text{ odd } \,. 
    \end{cases}
\end{equation}
The decomposition in equation \eqref{eqn:sodecomp}, provides a branching rule for the each representation of $\mathfrak{so}(N)$ under the decomposition. The branching rule for the vector representation can be read off straightforwardly from the BD-partition in equation \eqref{eqn:BDpart}, and it is:
\begin{equation}\label{eqn:sovectdecomp}
    \bm{N} \quad \longrightarrow \quad \bigoplus_{i=1}^{N} V_i \otimes R_i \,.
\end{equation}
As in equation \eqref{eqn:sufunddecomp}, $V_i$ is the $i$-dimensional representation of $\mathfrak{su}(2)_X$; however, in contrast, here $R_i$ is the fundamental representation of $\mathfrak{usp}(m_i)$ if $i$ is even, and the vector representation of $\mathfrak{so}(m_i)$ when $i$ is odd. The Dynkin embedding index for the $\mathfrak{su}(2)_X$ factor and for each $\mathfrak{j}(m_i)$ factor is determined via the branching, and we find
\begin{equation}\label{eqn:soDI}
    I_X = \frac{1}{12} \sum_{i=1}^N m_i i (i^2 - 1) \,, \qquad 
    I_i = \begin{cases} 
    &i/2 \qquad \text{if } i \text{ even}, \\
    &i \quad\qquad \text{if } i \text{ odd}.
    \end{cases}
\end{equation}

To determine the Nambu--Goldstone modes that arise when Higgsing via an arbitrary D-partition we need to understand the branching of the adjoint representation of $\mathfrak{so}(N)$ under the decomposition in equation \eqref{eqn:sodecomp}. This is straightforward via recalling that the adjoint representation of $\mathfrak{so}(N)$ is the anti-symmetric component of the tensor product of two vectors, 
\begin{equation}
    \textbf{adj} = \operatorname{ASym}(\bm{N} \otimes \bm{N}) \,,
\end{equation}
combined with the decomposition in equation \eqref{eqn:sufunddecomp}. In particular, we find
\begin{equation}\label{eqn:adjdecompSO}
  \begin{aligned}
    \textbf{adj} &\rightarrow \bigoplus_{\substack{i,j=1 \\  j > i}}^{N} (V_i \otimes V_j, R_i \otimes R_j) \, \oplus\, \bigoplus_{i=1}^N  (\operatorname{ASym}(V_i \otimes V_i), \operatorname{Sym}(R_i \otimes R_i)) \\ &\qquad\qquad\qquad \oplus \bigoplus_{i=1}^N (\operatorname{Sym}(V_i \otimes V_i), \operatorname{ASym}(R_i \otimes R_i)) \,. 
  \end{aligned}
\end{equation}
To expand fully in terms of irreducible representations, we first give the symmetric and anti-symmetric irreps appearing in the tensor products of $\mathfrak{su}(2)$ representations. In particular:
\begin{equation}\label{eqn:su2ASS}
    \begin{gathered}
        \operatorname{ASym}(V_{2i} \otimes V_{2i}) = \bigoplus_{k=1}^i V_{4i + 1 - 4k} \,, \quad \operatorname{ASym}(V_{2i-1} \otimes V_{2i-1}) = \bigoplus_{k=1}^{i-1} V_{4i - 1 - 4k} \,, \\
        \operatorname{Sym}(V_{2i} \otimes V_{2i}) = \bigoplus_{k=1}^i V_{4i + 3 - 4k} \,, \quad \operatorname{Sym}(V_{2i-1} \otimes V_{2i-1}) = \bigoplus_{k=1}^{i} V_{4i + 1 - 4k} \,.
    \end{gathered}
\end{equation}
When $i$ is even, then $R_i$ is the fundamental representation of $\mathfrak{usp}(m_i)$, and the symmetric and anti-symmetric tensor products of this fundamental representation decomposes into irreducible representations as
\begin{equation}\label{eqn:uspASS}
    \operatorname{ASym}(R_i \otimes R_i) = \bm{A^2} \oplus \bm{1} \,, \quad \operatorname{Sym}(R_i \otimes R_i) = \textbf{adj} \,.
\end{equation}
Similarly, when $i$ is odd and thus $R_i$ is the vector representation of $\mathfrak{so}(m_i)$, then the tensor products decompose into irreps as follows:
\begin{equation}\label{eqn:soASS}
    \operatorname{ASym}(R_i \otimes R_i) = \textbf{adj} \,, \quad \operatorname{Sym}(R_i \otimes R_i) = \bm{S^2} \oplus \bm{1} \,.
\end{equation}

Therefore, we can determine the coefficients appearing in the anomaly polynomial of the Nambu--Goldstone modes. We find
\begin{equation}\label{eqn:soHs}
  \begin{aligned}
    H &= \frac{1}{2} \bigg( \bigg[ \frac{1}{2}\sum_{i, j = 1}^{N} m_i m_j (ij - \operatorname{min}(i,j)) \bigg] - \frac{N}{2} + \frac{1}{2} \sum_{\substack{i = 1 \\ i \text{ odd}}}^N m_i \bigg) \,, \\
    H_v &= \frac{1}{2} \bigg[ \frac{1}{6}\sum_{i, j = 1}^{N} m_i m_j (ij - \operatorname{min}(i,j))(i(i-2) + j(j-2) + 2\operatorname{min}(i,j) - 1) \bigg] \\ &\qquad\qquad + \frac{1}{12}\sum_{i = 1}^N (i + 6i^2 - 4i^3)m_i - \frac{1}{12} \sum_{\substack{i = 1 \\ i \text{ odd}}}^N 3 m_i \,, \\
    H_i &= \begin{cases}
            \displaystyle \left[ 2\sum_{j=1}^N (ij - \operatorname{min}(i,j))m_j \right] - 4(i-1) &\qquad \text{ if $i$ odd} \\[1.6em]
            \displaystyle \frac{1}{2} \bigg[ 2\sum_{j=1}^N (ij - \operatorname{min}(i,j))m_j\bigg] - 2i  &\qquad \text{ if $i$ even} \,.
    \end{cases}
  \end{aligned}
\end{equation}
Here, the terms written in square brackets are the $H$, $H_v$, and $H_i$ for the case where $G = SU(N)$, as appear in equation \eqref{eqn:SUNGHs}. To determine the $H_i$, it was necessary to know the trace relations for $\mathfrak{so}$ and $\mathfrak{usp}$ algebras. In particular, we have used the following:
\begin{equation}\label{eqn:sospTR}
  \begin{aligned}
    \mathfrak{so}(m_i) \,: &\quad A_\textbf{vector} = 1 \,, & &\, A_\textbf{anti-sym} = h^\vee_{\mathfrak{so}(m_i)} \,, & &\, A_\textbf{sym} = h^\vee_{\mathfrak{so}(m_i)} + 4 \,, \\
    \mathfrak{usp}(m_i) \,: &\quad A_\textbf{fund} = \frac{1}{2} \,, & &\, A_\textbf{anti-sym} = h^\vee_{\mathfrak{usp}(m_i)} - 2 \,, & &\, A_\textbf{sym} = h^\vee_{\mathfrak{usp}(m_i)}\,.
  \end{aligned}
\end{equation}

\subsection{\texorpdfstring{$G = USp(N)$}{G = USp(N)}}\label{sec:USp}

Nilpotent orbits of $USp(N)$ are in one-to-one correspondence with C-partitions of $N$. A C-partition of $N$ is 
\begin{equation}\label{eqn:Cpart}
    [1^{m_1}, 2^{m_2}, \cdots, N^{m_N}] \quad \text{ such that } \quad \sum_{i=1}^N i m_i = N \quad \text{ and } \quad i \text{ odd } \,\, \Longrightarrow \,\, m_i \text{ even } \,.
\end{equation}
Notice that the constraint in equation \eqref{eqn:Cpart} that the odd entries in the partition appear with even multiplicity implies that $N$ must be an even integer. As we are by now familiar, nilpotent orbits of $\mathfrak{g}$ are in one-to-one correspondence with conjugacy classes of $\mathfrak{su}(2)$ embeddings in $\mathfrak{g}$; the centralizer of this embedding is $\mathfrak{f}$, as in equation \eqref{eqn:su2Xembed}. For the C-partition in equation \eqref{eqn:Cpart}, this is
\begin{equation}
    \mathfrak{f} = \bigoplus_{i=1}^N \mathfrak{j}_i \qquad \text{ where } \qquad \mathfrak{j}_i = \begin{cases}
        \mathfrak{usp}(m_i) \quad &\text{ if } i \text{ odd } \\
        \mathfrak{so}(m_i) \quad &\text{ if } i \text{ even } \,,
    \end{cases}
\end{equation}
and the branching of the fundamental representation of $\mathfrak{usp}(N)$ under this decomposition is
\begin{equation}\label{eqn:uspfunddecomp}
    \bm{N} \rightarrow \bigoplus_{i=1}^N V_i \otimes R_i \,.
\end{equation}
Again, $R_i$ is either the fundamental representation of $\mathfrak{usp}(m_i)$ or the vector representation of $\mathfrak{so}(m_i)$, depending on whether $i$ is odd or even, respectively. From the branching rule in equation \eqref{eqn:uspfunddecomp}, we determine that the Dynkin embedding indices for each simple factor in the decomposition in equation \eqref{eqn:su2Xembed} are
\begin{equation}\label{eqn:uspDI}
    I_X = \frac{1}{6} \sum_{i=1}^N m_i i (i^2 - 1)  \,, \qquad I_i = \begin{cases} 2i \qquad &\text{if } i \text{ even } \\
    i \qquad &\text{if } i \text{ odd } \,.
    \end{cases}
\end{equation}

To determine the spectrum of Nambu--Goldstone modes, we need the branching rules for the adjoint representation under the embedding. As the adjoint of $\mathfrak{usp}(N)$ is the symmetric component of the tensor product of the fundamental, 
\begin{equation}
    \textbf{adj} = \operatorname{Sym}(\bm{N} \otimes \bm{N}) \,,
\end{equation}
we can determine the branching from equation \eqref{eqn:uspfunddecomp}. We find
\begin{equation}\label{eqn:adjdecompUSP}
  \begin{aligned}
    \textbf{adj} &\rightarrow \bigoplus_{\substack{i,j=1 \\  j > i}}^{N} (V_i \otimes V_j, R_i \otimes R_j) \, \oplus\, \bigoplus_{i=1}^N  (\operatorname{ASym}(V_i \otimes V_i), \operatorname{ASym}(R_i \otimes R_i)) \\ &\qquad\qquad\qquad \oplus \bigoplus_{i=1}^N (\operatorname{Sym}(V_i \otimes V_i), \operatorname{Sym}(R_i \otimes R_i)) \,. 
  \end{aligned}
\end{equation}
This can again be expanded as a direct sum of irreducible representations using equations \eqref{eqn:su2ASS}, \eqref{eqn:uspASS}, and \eqref{eqn:soASS}. Now it is straightforward to determine the coefficients appearing in the anomaly polynomial of the Nambu--Goldstone modes. The final result is
\begin{equation}\label{eqn:uspHs}
  \begin{aligned}
    H &= \frac{1}{2} \bigg( \bigg[ \frac{1}{2}\sum_{i, j = 1}^{N} m_i m_j (ij - \operatorname{min}(i,j)) \bigg] + \frac{N}{2} - \frac{1}{2} \sum_{\substack{i = 1 \\ i \text{ odd}}}^N m_i \bigg) \,, \\
    H_v &= \frac{1}{2} \bigg[ \frac{1}{6}\sum_{i, j = 1}^{N} m_i m_j (ij - \operatorname{min}(i,j))(i(i-2) + j(j-2) + 2\operatorname{min}(i,j) - 1) \bigg] \\ &\qquad\qquad - \frac{1}{12}\sum_{i = 1}^N (i + 6i^2 - 4i^3)m_i + \frac{1}{12} \sum_{\substack{i = 1 \\ i \text{ odd}}}^N 3 m_i \,, \\
    H_i &= \begin{cases}
            \displaystyle \frac{1}{2} \bigg[ 2\sum_{j=1}^N (ij - \operatorname{min}(i,j))m_j \bigg]+ 2(i-1) &\qquad \text{ if $i$ odd} \\[1.6em]
            \bigg[2\displaystyle \sum_{j=1}^N (ij - \operatorname{min}(i,j))m_j\bigg] + 4i  &\qquad \text{ if $i$ even} \,.
    \end{cases}
  \end{aligned}
\end{equation}
Once again, the terms in square brackets are the $H$, $H_v$, and $H_i$ that appear in the Nambu--Goldstone analysis of $SU(N)$, collected in equation \eqref{eqn:SUNGHs}.
For the determination of $H_i$, we again had to make use of the trace identities appearing in equation \eqref{eqn:sospTR}.

\section{4d \texorpdfstring{\boldmath{$\mathcal{N}=2$}}{N=2} SCFTs of class \texorpdfstring{\boldmath{$\mathcal{S}$}}{S}}\label{sec:Fieldtheory}

A vast landscape of strongly-coupled 4d $\mathcal{N}=2$ SCFTs
arise via the class $\mathcal{S}$ construction
\cite{Gaiotto:2009we,Gaiotto:2009hg}. The class $\mathcal{S}$ SCFTs can be constructed by compactifying a 6d
$(2,0)$ SCFT of type $\mathfrak{j}$, where $\mathfrak{j}$ is a simple and
simply-laced Lie algebra, on an $n$-punctured
genus-$g$ Riemann surface. Each puncture is associated to a choice of an ($\mathcal{N}=2$)-preserving
codimension-two defect in the 6d $(2,0)$ SCFT. A class $\mathcal{S}$ theory can then be denoted as
\begin{equation}\label{eqn:classSgen}
    \mathcal{S}_\mathfrak{j}\langle C_{g,n} \rangle \{ O_1, \cdots, O_n \} \,,
\end{equation}
where $\mathcal{S}_\mathfrak{j}$ is the 6d $(2,0)$ SCFT of type $\mathfrak{j}$,
the $\langle C_{g,n} \rangle$ indicates the eight-supercharge-preserving
twisted compactification on an $n$-punctured genus-$g$ Riemann surface, and each
$O_i$ specifies one of the $n$ punctures. The possible puncture data $O_i$ has
been extensively studied, see, for example, \cite{Chacaltana:2010ks,Chacaltana:2011ze,Chacaltana:2012zy,Chacaltana:2012ch,Chacaltana:2013oka,Chacaltana:2014jba,Chacaltana:2015bna,Chacaltana:2016shw,Chacaltana:2017boe,Chacaltana:2018vhp,Xie:2012hs,Xie:2013jc,Wang:2015mra,Wang:2018gvb}. 

The most straightforward class of punctures are the \emph{untwisted regular} punctures; these are in one-to-one correspondence with
the nilpotent orbits of $\mathfrak{j}$. The second class of punctures that we consider are the \emph{twisted regular} punctures. When $\mathfrak{j}$ admits an outer-automorphism $o$, we can twist by this outer-automorphism around any puncture on the Riemann surface. Let $\mathfrak{g}^\vee$ denote the invariant subalgebra of $\mathfrak{j}$ under the action of $o$, and $\mathfrak{g}$ be the Langlands dual of $\mathfrak{g}^\vee$. Twisted regular punctures are in one-to-one correspondence with nilpotent orbits of $\mathfrak{g}$. In fact, twisted punctures come in groups; this is necessary to globally cancel off the monodromy introduced by the outer-automorphism twist. Since the only outer-automorphism groups of simple Lie algebras are $\mathbb{Z}_2$ and $S_3$, the twisted punctures must come either in pairs or triples. In this paper, we consider only the cases where $\mathfrak{g}$ is a classical algebra, and thus only the former is possible. 

For a classical Lie algebra $\mathfrak{g}$, each nilpotent orbit can be associated to an integer partition. For $\mathfrak{g} = \mathfrak{su}(n),\, \mathfrak{usp}(2n),\, \mathfrak{so}(2n+1)$, nilpotent orbits are in one-to-one correspondence with partitions of $n$, C-partitions of $2n$, and B-partitions of $2n+1$, respectively.\footnote{A B-partition or D-partition of $N$ is an integer partition of $N$ such that every even element appears an even number of times; typically such a partition is called a B-partition when $N$ is odd and a D-partition when $N$ is even. A C-partition of $N$ is an integer partition of $N$ such that every odd element appears an even number of times. A D-partition is very even if it consists of only even elements. For ease of notation, we sometimes refer to an integer partition of $N$ as an A-partition of $N$.} For $\mathfrak{g}=\mathfrak{so}(2n)$, nilpotent orbits are associated to D-partitions of $2n$; this is one-to-one except when the D-partition is very even, in which case the D-partition is associated to two distinct nilpotent orbits. We do not stress this subtlety in this paper, as the
conformal anomalies depend on the D-partition itself, but not the underlying nilpotent orbits. The distinction between the two nilpotent orbits is generically captured by the higher R-charge spectrum of the theory, as
explored in \cite{Distler:2022nsn,Distler:2022yse}.

For any simple Lie algebra, the set of nilpotent orbits admits a partial ordering $\prec$. Viewing the nilpotent orbits in terms of their associated integer partitions, the partial ordering is given by the \emph{dominance ordering}, which is defined as the following. Let $P$ and $P'$ be integer partitions of $N$. We can represent them as length-$N$ sequences
\begin{equation}
    P = [p_1, \cdots, p_N] \,, \qquad P' = [p'_1, \cdots, p'_N] \,,
\end{equation}
where $p_i$ and $p_i'$ are non-negative integers in weakly decreasing order.
Then, the dominance ordering is given by the following inequality:
\begin{equation}
    P' \prec P \quad \Longleftrightarrow \quad \sum_{i=j}^N (p_i' - p_i) \geq 0 \,\, \text{ for } \,\, 1 \leq j \leq N \,.
\end{equation}

Under this dominance ordering, there exists a unique maximal nilpotent orbit $O_\text{full}$ such that $O \prec O_\text{full}$ for
all other nilpotent orbits $O$. We refer to the puncture associated to $O_\text{full}$ as the \emph{full puncture}. As we will discuss shortly, all class $\mathcal{S}$ theories with regular punctures, untwisted and twisted, can be obtained from Higgs branch renormalization group flows from a class $\mathcal{S}$ theory with only full punctures. A puncture $O \prec O_\text{full}$ is said to be \emph{partially closed}, unless if $O$ is the smallest nilpotent orbit under the ordering $\prec$, for which we call it \emph{fully closed}. 

Class $\mathcal{S}$ theories with only regular punctures always give rise to 4d $\mathcal{N}=2$ SCFTs where the scaling dimensions of all of the Coulomb branch operators are integral. However, there is a broad and interesting class of strongly-coupled 4d $\mathcal{N}=2$ SCFTs, known as the Argyres--Douglas
theories \cite{Argyres:1995jj}, where the Coulomb branch scaling dimensions can be rational. Such 4d SCFTs can arise from the class $\mathcal{S}$ perspective via the inclusion of irregular punctures
\cite{Cecotti:2012jx,Xie:2012hs,Cecotti:2013lda,Wang:2015mra}. The presence of irregular punctures is highly restrictive -- the Riemann surface is forced to be a sphere, and there can only be at most two punctures: an irregular puncture and a regular puncture. The punctures must be either both untwisted or both twisted; hence, we refer to the theory itself as untwisted or twisted. We have discussed that a regular puncture is associated to a nilpotent orbit of $\mathfrak{g}$,\footnote{We abuse notation slightly by writing the untwisted punctures as punctures twisted by the trivial outer-automorphism, in which case $\mathfrak{g} = \mathfrak{j}$.} and it remains for us to describe how an irregular puncture is parametrized.

In this paper, we consider ``irregular punctures of regular semi-simple type''
\cite{Wang:2015mra,Wang:2018gvb}. Such punctures can be parametrized by a pair
of integers $(b, p)$, where $p$ is positive and $b$ can take a small set of
positive values depending on $\mathfrak{j}$ and $o$.\footnote{There are (at
least) two standard conventions in the literature for the irregular punctures
of regular semi-simple type that we consider in this paper. Either, the
puncture is parametrized by $(b, p)$ where $p > 0$, or else by $(b, k)$, where
$k > -b$. We will use $p$ exclusively throughout this paper; to convert to the
alternative convention, we note that $k = p - b$ in all cases.} To understand the possible punctures, it is useful to consider the compactification of the non-Lagrangian 6d $(2,0)$ SCFT on a circle that is transverse to the Riemann surface. This is believed to lead to 5d $\mathcal{N}=2$ super Yang--Mills on $\mathbb{R}^{1,2} \times C_{g,n}$; this is a Lagrangian field theory involving a scalar field $\Phi$, known as the Higgs field, and $\Phi$ has poles at the location of the punctures on $C_{g,n}$. Let $z$ be a local coordinate on the Riemann surface, and $z = 0$ be the location of a puncture. We can write the Higgs field around the puncture as
\begin{equation}\label{eqn:Higgs}
    \Phi_z = \sum_{p \geq q \geq 0} \frac{T_q}{z^{1 + q/b}} \,,
\end{equation}
where $b$ is a positive integer, $p$ an integer, and $T_q$ a semi-simple element of $\mathfrak{j}$. The Higgs field furthermore needs to satisfy certain conditions (see \cite{Wang:2018gvb}) to be well-defined on $C_{g,n}$. Irregular punctures occur when the order of the pole in equation \eqref{eqn:Higgs} is larger than one; the irregular punctures of regular semi-simple type further require that the numerator $T_{p}$ is a regular semi-simple element of $\mathfrak{j}$. The conditions for $\Phi_z$ to be well-defined on $C_{g,n}$ then lead to constraints on the possible values of $b$ and $p$, as explained in detail in \cite{Wang:2015mra,Wang:2018gvb}. We list the
possible values of $b$ and $p$ for $\mathfrak{g}$ a classical Lie algebra in Table
\ref{tbl:irregs}. 

Since class $\mathcal{S}$ theories involving irregular punctures are so constrained, we introduce the following notation for each
theory, simplifying that in equation \eqref{eqn:classSgen}:
\begin{equation}\label{eqn:twgen}
    \big(\mathfrak{j}, o, \mathfrak{g}, b, p, O \big) \quad \rightarrow \quad \mathcal{D}_p^{\, b}(G, O) \,,
\end{equation}
where we simply write the automorphism group itself for $o$, and $O$ is a
nilpotent orbit of $\mathfrak{g}$. When $O = O_\text{full}$ is the full puncture, we often drop the nilpotent orbit entirely:\footnote{When writing $\mathcal{D}_{p}^{\,b}(G, O)$,
the Lie algebra $\mathfrak{g}$ is usually written in group notation, e.g.,
$SU(N)$ instead of $\mathfrak{su}(N)$. We use $\mathfrak{g}$ and $G$ (and similarly $\mathfrak{j}$ and $J$) interchangeably throughout this paper.}
\begin{equation}
    \big(\mathfrak{j}, o, \mathfrak{g}, b, p, O_\text{full} \big) \quad \rightarrow \quad \mathcal{D}_p^{\, b}(G) \,.
\end{equation}

\begin{table}[p]
    \centering
    \renewcommand{\arraystretch}{1.2}
    \begin{threeparttable}
        \begin{tabular}{cccccc}
          \toprule
          $\mathfrak{j}$ & $o$ & $\mathfrak{g}$ & $b$ & $p$ & $O$ \\\midrule
          \multirow{2}{*}{$\mathfrak{su}(n)$} & \multirow{2}{*}{$1$} & \multirow{2}{*}{$\mathfrak{su}(n)$} & $n$ & $\mathbb{N}$ & \multirow{2}{*}{A-part.~of $n$} \\
           & & & $n-1$ & $\mathbb{N}$ \\\midrule
          \multirow{2}{*}{$\mathfrak{su}(2n + 1)$} & \multirow{2}{*}{$\mathbb{Z}_2$} & \multirow{2}{*}{$\mathfrak{usp}'(2n)$} & $4n+2$ & $\mathbb{N}_\text{odd}$ & \multirow{2}{*}{C-part.~of $2n$} \\
           & & & $2n$ & $\mathbb{N}$ \\\midrule
          \multirow{2}{*}{$\mathfrak{su}(2n)$} & \multirow{2}{*}{$\mathbb{Z}_2$} & \multirow{2}{*}{$\mathfrak{so}(2n+1)$} & $4n-2$ & $\mathbb{N}_\text{odd}$ & \multirow{2}{*}{B-part.~of $2n+1$} \\
           & & & $2n$ & $\mathbb{N}$ \\\midrule
          \multirow{2}{*}{$\mathfrak{so}(2n)$} & \multirow{2}{*}{$1$} & \multirow{2}{*}{$\mathfrak{so}(2n)$} & $2n-2$ & $\mathbb{N}$ & \multirow{2}{*}{D-part.~of $2n$} \\
           & & & $n$ & $\mathbb{N}$ & \\\midrule
          \multirow{2}{*}{$\mathfrak{so}(2n + 2)$} & \multirow{2}{*}{$\mathbb{Z}_2$} & \multirow{2}{*}{$\mathfrak{usp}(2n)$} & $2n+2$ & $\mathbb{N}_\text{odd}$ & \multirow{2}{*}{C-part.~of $2n$} \\
           & & & $2n$ & $\mathbb{N}$ \\
          \bottomrule
        \end{tabular}
    \end{threeparttable}
    \caption{For each pair $(\mathfrak{j}, o)$, we list the Langlands dual of the invariant subalgebra $\mathfrak{g}^\vee\subset\mathfrak{j}$, the pairs $(b, p)$ parametrizing an irregular puncture of regular semi-simple type, and the type of integer partition describing the nilpotent orbit of $\mathfrak{g}$ associated to a regular puncture, $O$. For each pair $(\mathfrak{j}, o)$, the first-listed value of $b$ is $b = h_t$, the twisted Coxeter number, as listed in Table \ref{tbl:grp}. By convention, the $\mathfrak{usp}(2n)$ obtained from the $\mathbb{Z}_2$ outer-automorphism of $\mathfrak{su}(2n+1)$ is denoted as $\mathfrak{usp}'(2n)$.
    }
    \label{tbl:irregs}
    \vspace{0.3cm}

    \renewcommand{\arraystretch}{1.2}
    \begin{threeparttable}
        \begin{tabular}{cccc}
          \toprule
          $\mathfrak{j}$ & $o$ & $h_t$ & $\operatorname{Cas}_t(J,o)$ \\\midrule
          $\mathfrak{su}(2n + 1)$ & $\mathbb{Z}_2$ & $4n+2$ & $\left\{3, 5, \cdots, 2n+1 \right\}$ \\
          $\mathfrak{su}(2n)$ & $\mathbb{Z}_2$ & $4n-2$ & $\left\{3, 5, \cdots, 2n-1 \right\}$ \\
          $\mathfrak{so}(2n + 2)$ & $\mathbb{Z}_2$ & $2n+2$ & $\left\{ n+1 \right\}$ \\
          \bottomrule
        \end{tabular}
    \end{threeparttable}\quad
    \begin{threeparttable}
        \begin{tabular}{cc}
          \toprule
          $\mathfrak{g}$ & $\operatorname{Cas}(G)$ \\\midrule
          $\mathfrak{su}(n)$ & $\left\{2, 3, \cdots, n \right\}$ \\
          $\mathfrak{so}(2n + 1)$ & $\left\{2, 4, \cdots, 2n \right\}$ \\
          $\mathfrak{so}(2n)$ & $\left\{2, 4, \cdots, 2n-2, n \right\}$ \\
          $\mathfrak{usp}(2n)$ & $\left\{2, 4, \cdots, 2n \right\}$ \\
          \bottomrule
        \end{tabular}
    \end{threeparttable}
    \caption{For each pair $(\mathfrak{j}, o)$, with $o$ non-trivial, we list the twisted Coxeter number, $h_t$, and the twisted Casimir degrees, $\operatorname{Cas}_t(J,o)$. When $o = 1$, we have $h_t = h_\mathfrak{j}$ and $\operatorname{Cas}_t(J,o) = \varnothing$. For convenience, we also list the degrees of the Casimir invariants for each simple classical Lie algebra, $\mathfrak{g}$.
    }
    \label{tbl:grp}
\end{table}

Now that we have enumerated the class $\mathcal{S}$ theories that we are interested in, we would like to be able to determine some more of their physical properties, such as their conformal anomalies and flavor central charges. 

First, let us determine the central charges of a class $\mathcal{S}$ theory
with $n$ regular untwisted maximal punctures, and $m$ regular twisted maximal
punctures. The central charges factorize into a contribution from the genus,
and a sum of contributions of the punctures. We have 
\begin{equation}
  \begin{aligned}
    a &= \frac{1}{24}(g-1)(8 d_\mathfrak{j} h_\mathfrak{j}^\vee + 5 r_\mathfrak{j}) + n \delta a(O_\text{full}) + m \delta a(\widetilde{O}_\text{full}) \,,\\
    c &= \frac{1}{6}(g-1)(2 d_\mathfrak{j} h_\mathfrak{j}^\vee + r_\mathfrak{j}) + n \delta c(O_\text{full}) + m \delta c(\widetilde{O}_\text{full}) \,,
  \end{aligned}
\end{equation}
where $d_\mathfrak{j}$, $h_\mathfrak{j}^\vee$, and $r_\mathfrak{j}$ are, respectively, the dimension, dual Coxeter number, and the rank of $\mathfrak{j}$.
Furthermore, $\delta a(O_\text{full})$, $\delta c(O_\text{full})$ are the respective
contributions from the untwisted full punctures, and similarly $\delta
a(\widetilde{O}_\text{full})$, $\delta c(\widetilde{O}_\text{full})$ are the
contributions from each of the twisted full punctures. These contributions can each 
be worked out straightforwardly following \cite{Chacaltana:2012zy}. 

We find the contributions to the central charges $a$ and $c$ to be
\begin{equation}
  \begin{aligned}
    \delta a(O_\text{full}) &= \frac{1}{48} \left( d_{\mathfrak{j}}(8h^\vee_{\mathfrak{j}} - 5) + 5r_{\mathfrak{j}}) \right) \,, &\qquad
    \delta c(O_\text{full}) &= \frac{1}{12} \left( d_{\mathfrak{j}}(2h^\vee_{\mathfrak{j}} - 1) + r_{\mathfrak{j}}) \right) \,, \\
    \delta a(\widetilde{O}_\text{full}) &= \frac{1}{48} \left( 8d_{\mathfrak{j}}h^\vee_{\mathfrak{j}} - 5d_{\mathfrak{g}} + 5r_{\mathfrak{j}} \right) \,, &\qquad
    \delta c(\widetilde{O}_\text{full}) &= \frac{1}{12} \left( 2d_{\mathfrak{j}}h^\vee_{\mathfrak{j}} - d_{\mathfrak{g}} + r_{\mathfrak{j}}\right) \,.
  \end{aligned}
\end{equation}
The flavor symmetries and flavor central changes also factorize over the
punctures:\footnote{The actual flavor symmetry can be larger than this manifest
symmetry, however that does not cause us undue concern as the nilpotent
Higgsing we consider in this paper will be with respect to the manifest flavor
symmetries of the maximal punctures.}
\begin{equation}
	\mathfrak{f} = \mathfrak{j}_{k_\mathfrak{j}}^{\oplus n} \oplus \mathfrak{g}_{\widetilde{k}_\mathfrak{g}}^{\oplus m} \,,
\end{equation}
where $\mathfrak{j}$ is the 6d $(2,0)$ type, and $\mathfrak{g}$ is the
Langlands dual of the invariant subalgebra of $\mathfrak{j}$ under $o$, as
usual. The flavor central charges are
\begin{equation}
    k_\mathfrak{j} = 2h_{\mathfrak{j}}^\vee \,, \qquad \widetilde{k}_\mathfrak{g}=2h_{\mathfrak{g}}^\vee \,.
\end{equation}

We now turn to the anomaly polynomials of the theories with irregular
punctures.\footnote{The results appearing below are an amalgamation of results that have appeared throughout the literature involving irregular punctures in class $\mathcal{S}$, with varying conventions and normalizations. See, for example, \cite{Wang:2015mra,Wang:2018gvb,Cecotti:2012jx,Cecotti:2013lda,Carta:2022spy,Carta:2021whq,Giacomelli:2020ryy,Giacomelli:2017ckh}.}  To determine the conformal anomalies of the theories
$\mathcal{D}_p^{\, b}(G,O)$ as in equation \eqref{eqn:twgen}, it suffices to
understand the anomalies of the theory $\mathcal{D}_p^{\, b}(G)$,
and then the anomalies after partial closure of the full puncture follow from
the procedure described in Section \ref{sec:nilphiggsing}.

We first consider the case where there are no twist lines, i.e.,
\begin{align}
    o = 1\,,\quad \mathfrak{j} = \mathfrak{g} \,,
\end{align}
in equation \eqref{eqn:twgen}. In this case, when $b = h_{\mathfrak{g}}$, we often drop the explicit $b$ and simply write $\mathcal{D}_p(G)$. The full puncture contributes a factor of $\mathfrak{g}$ to the flavor algebra
of the theory, and the level of this flavor symmetry is\footnote{There are two common normalizations for the flavor central charge in the literature: ours and $\tfrac{1}{2}$ours. To create a reference point: our normalization is such that the rank one Minahan--Nemeschansky theory with flavor symmetry $\mathfrak{e}_8$ \cite{Minahan:1996cj} has flavor central charge $k_{\mathfrak{e}_8} = 12$.}
\begin{equation}
    k_\mathfrak{g} = 2\left ( h_\mathfrak{g}^\vee - \frac{b}{p} \right) \,,
\end{equation}
where $h_\mathfrak{g}^\vee$ is again the dual Coxeter number of $\mathfrak{g}$.
The central charge $c$ can be written in terms of this flavor central charge:
\begin{equation}\label{eqn:cUTW}
    12c = \frac{k_\mathfrak{g}d_{\mathfrak{g}}}{2h_\mathfrak{g}^\vee - k_\mathfrak{g}} - f \,.
\end{equation}
The quantity $f$ is the number of mass parameters associated to the irregular
puncture; that is, the full flavor symmetry of the class $\mathcal{S}$ theory
contains
\begin{equation}
    \mathfrak{g} \oplus \mathfrak{u}(1)^{\oplus f} \,,
\end{equation}
as a subalgebra. We discuss how to determine $f$ shortly. To
determine the third conformal anomaly we can use the Shapere--Tachikawa formula \cite{Shapere:2008zf}:
\begin{equation}
    4(2a - c) = \sum_u \left( 2\Delta(u) - 1 \right) \,,
\end{equation}
where the sum runs over the Coulomb branch operators, $u$, of the theory, and
where $\Delta(u)$ is the scaling dimension of $u$. To apply this formula we
need to know the set of $\Delta(u)$. For $\mathcal{D}_p(G)$, the set of Coulomb branch scaling dimensions is
\begin{equation}\label{eqn:CBuntw}
  \left\{ \Delta(u) \right\} =  \left\{j - \frac{b}{p}s > 1\,\,\bigg|\,\, j \in \operatorname{Cas}(G)\,,\, s \geq 1 \right\} \,,
\end{equation}
where $\operatorname{Cas}(G)$ is the set of Casimir invariants of $\mathfrak{g}$. 

We now turn to the class $\mathcal{S}$ theories with a twisted irregular puncture. To understand the anomalies for an arbitrary Higgsing, we need to first determine the anomalies of $\mathcal{D}_p^{\, b}(G)$, where
\begin{equation}
    o \neq 1\,,\quad \mathfrak{j} \neq \mathfrak{g} \,,
\end{equation} 
in which case there must be a $\mathbb{Z}_2$ twist-line. For these theories, the flavor algebra associated to the full regular puncture is $\mathfrak{g}$, and the flavor central charge is given by the unified formula
\begin{equation}\label{eq:flavorDpG}
    k_{\mathfrak{g}} = 2\left( h_\mathfrak{g}^\vee - \frac{1}{m} \frac{b}{p} \right) \,,
\end{equation}
where $m$ depends on the pair $(\mathfrak{j}, o)$.\footnote{For $(\mathfrak{so}(2n+2), \mathbb{Z}_2)$ and $(\mathfrak{su}(2n), \mathbb{Z}_2)$ we have $m = 2$, for $(\mathfrak{su}(2n+1), \mathbb{Z}_2)$ we have $m = 4$, and for all other combinations $m = 1$.} For each of these families of theories, the central charge $c$ is given by the obvious generalization of equation \eqref{eqn:cUTW}:
\begin{equation}\label{eqn:cONk}
    12c = \frac{k_\mathfrak{g}d_{\mathfrak{g}}}{2h_\mathfrak{g}^\vee - k_\mathfrak{g}} - f \,,
\end{equation}
where $f$ is again the number of mass parameters. To determine the central charge $a$, we again use the Shapere--Tachikawa formula, for which we need to know the spectrum of Coulomb branch operators of the theories $\mathcal{D}_p^{\, b}(G)$ involving twisted punctures. This is again a natural generalization of equation \eqref{eqn:CBuntw} for the untwisted theories. We find that the spectrum of Coulomb branch operators for all theories $\mathcal{D}_p^{\,b}(G)$, i.e., given by the data $(\mathfrak{j}, o, \mathfrak{g}, b, p, O_\text{full})$, with $\mathfrak{g}$ a classical Lie algebra is
\begin{equation}\label{eqn:CBtw}
  \left\{j - \frac{b}{p}s > 1\,\,\bigg|\,\, j \in \operatorname{Cas}(G)\,,\, s \geq 1 \right\} \, \sqcup \left\{j_t - \frac{b}{p} \frac{2s-1}{2} > 1\,\,\bigg|\,\, j_t \in \operatorname{Cas}_t(J, o)\,,\, s \geq 1 \right\} \,.
\end{equation}
Here, $\operatorname{Cas}(G)$ is the degrees of the fundamental Casimir invariants of the algebra $\mathfrak{g}$, and $\operatorname{Cas}_t(J, o)$ are the degrees of the ``twisted Casimir invariants'' under the action of the outer-automorphism $o$. The latter are listed in Table \ref{tbl:grp}.

Two other important physical properties are the number of mass parameters
contributed by the irregular puncture, which we have called $f$ above.  Now
that we have introduced the general formula for the spectrum of Coulomb branch
operators of $\mathcal{D}_p^{\,b}(G)$, it is straightforward to count this
quantity. The number of mass parameters is obtained as the cardinality of the
set in equation \eqref{eqn:CBtw} where the ``$> 1$'' is replaced with ``$=1$''.
The closed form expression typically depend in a somewhat involved way on
$\gcd$-like conditions, as they exist on when there are solutions to certain
equations over the integers. It is tedious to write these closed-form
expressions explicitly, see \cite{Giacomelli:2017ckh,Wang:2018gvb}. However, it
is clear that each element of $\operatorname{Cas}(G)$ and
$\operatorname{Cas}_t(J, o)$ can contribute at most one mass parameter;
therefore $f$ provides a subsubleading contribution to the central charge $c$
in equation \eqref{eqn:cONk}. As such, we do not expect the supergravity dual
solutions that we write down in this paper to be sensitive to $f$.

We have now determined the conformal anomalies for all class $\mathcal{S}$ theories where all the regular punctures are maximal. In the previous section, we showed how to determine the conformal anomalies of all class $\mathcal{S}$ theories, with arbitrary regular punctures, using the ``ultraviolet'' anomalies worked out here, together with the data of the nilpotent orbit which describes how the regular maximal puncture is partially closed.

\section{Holographic duals of Higgsed \texorpdfstring{\boldmath{$\mathcal{D}_p(ABCD)$}}{DpABCD}}\label{sec:holoDpG} 

In Section \ref{sec:Fieldtheory} we have discussed the various M-theoretic origins of the $\mathcal{D}_p^{\,b}(G)$ theories which were summarized in Table \ref{tbl:irregs}. In this section we will discuss their holographic duals. The key observation in this endeavour is to first understand the geometric origin of the parent 6d SCFT before realizing holographically the twist lines as codimension one topological defects of the parent theory. The former depends on a choice of $\mathbb{Z}_2$ orbifold of the internal space. The twist lines have two distinct realizations, either by a different choice of orbifold, or by including M5-branes fixed on top of the orbifold fixed plane. 

In the following subsection we discuss the geometric form of the solution that we require, before studying the explicit gravity duals. As we will discuss, the local form of the metric was classified in \cite{Lin:2004nb}. The global completion, in particular how to include the required orbifolding and the regularity and flux quantization, form the bulk of the section and generalize previous solutions within the classification of \cite{Lin:2004nb}. 

\subsection{Local \texorpdfstring{$\mathcal{N}=2$}{N=2} AdS\texorpdfstring{$_5$}{5} solutions of M-theory}

We are interested in identifying the $\mathcal{N}=2$ preserving AdS$_5$ solutions which are holographically dual to the $\D_p^{\,b}(G)$ theories. The local form of all $\mathcal{N}=2$ preserving AdS$_5$ solutions was classified in \cite{Lin:2004nb}, known as LLM. They showed that the most general form is determined in terms of a scalar potential satisfying the $SU(\infty)$ Toda equation. The solution takes the form
\begin{equation} \label{eq : LLM solution}
\begin{split}
 \dd s^2_{11} &= \kappa^{2/3} \me^{2\lambda} \bigg[ 4\dd s^2_{\text{AdS}_5} + y^2 \me^{-6\lambda} \dd s^2(S^2)+\dd s^2(\M_4)\bigg]\,, \\
\dd s^2(\M_4)&=4(1-y^2 \me^{-6\lambda}) +\frac{\me^{-6\lambda}}{1-y^2\me^{-6\lambda}}\Big( \dd y^2 +\me^D (\dd x_1^2 + \dd x_2^2)\Big) \,, \\
 G_4 &=\kappa \dd \vol(S^2) \wedge \bigg[ \mathcal{D}\chi\wedge \dd(y^3 e^{-6\lambda}) + y(1-y^2e^{-6\lambda})\dd v-\frac{1}{2}\partial_y \me^D \dd x_1 \wedge \dd x_2 \bigg]\,, \\
 \me^{-6\lambda} &= - \frac{\partial_y D}{y(1-y\partial_y D)}\,, \quad
\mathcal{D}\chi = \dd\chi + v\,, \quad 
v = \frac{1}{2} (\pd_{x_2} D \, \dd x_1 - \pd_{x_1} D \, \dd x_2) \,.
\end{split}
\end{equation}
With our conventions the $SU(\infty)$ Toda equation is
\begin{equation}
    \left(\partial_{x_1}^2+\partial_{x_2}^2\right) D+\partial_y^2 \me^{D}=0\, .
\end{equation}
We emphasize that this is a local solution, even after solving the Toda equation one must also impose additional constraints such that the solution is globally well-defined. In particular, one may replace the $S^2$ by $\mathbb{RP}^2$ without breaking supersymmetry. The metric has a $U(1)$ isometry generated by the Killing vector $\partial_{\chi}$ and an $SO(3)$ isometry generated by the two-sphere, these are the geometric duals of the R-symmetry of the field theory.

In general the Toda equation is difficult to solve and analytic solutions are somewhat scarce. We may bypass this difficulty by assuming that there is an additional $U(1)$ symmetry. To wit, let 
\begin{align}
    x_1 + i x_2 = r e^{i\beta} \,, 
\end{align}
with $\beta$ the coordinate of our new $U(1)$ action, and take the potential $D$ to be independent of it. With this additional isometry we may perform the following transformation, known as a Backl\"{u}nd transformation \cite{Ward:1990qt}
\begin{gather}
    r^2 e^D = \rho^2\,, \qquad y = \rho \partial_\rho V \equiv \dot{V}\,, \qquad \mathrm{log} r = \partial_\eta V \equiv V'\,, 
    \end{gather}
where the potential $V$ satisfies the 3d cylindrical Laplace equation
\begin{gather}
\frac{1}{\rho} \partial_\rho (\rho \partial_\rho V) + \partial_\eta^2 V = 0 \,.
\end{gather}

After performing the Backl\"und transformation the solution takes the form\footnote{There is a shift of the angular coordinate that we have suppressed here:
\begin{equation*}
\chi_T=\chi_E+\beta_{E}\, ,\quad  \beta_T=\beta_E\, ,
\end{equation*} 
with $T$ the Toda coordinate and $E$ the electrostatics (Backl\"und transformed) coordinate. The result is that the Killing vector dual to the R-symmetry is $R=\partial_{\chi}$ in both cases.} \cite{Gaiotto:2009gz}
\begin{equation}\label{eq:metric in terms of V in 11d}
\begin{aligned}
     \dd s_{11}^2 &= \kappa^{2/3} \bigg( \frac{\Vd\Dt}{2V''} \bigg)^{1/3} \bigg[4 \dd s^2_{\text{AdS}_5} + \frac{2V'' \Vd}{\Dt} \dd s^2(S^2) + \dd s_4^2 \bigg]\,, \\
     \dd s_4^2 &= \frac{2(2\Vd-\Vdd)}{\Vd\Dt} \mathcal{D}\beta^2 + \frac{2V''}{\Vd} \Big( \dd\rho^2 +\dd\eta^2 +  \frac{2\Vd }{2\Vd-\Vdd} \rho^2 \dd\chi^2 \Big)\,, \\
     C_3 &= 2 \kappa \left[ -  \frac{2\Vd^2 V''}{\Dt} \dd\chi + \Big( \frac{\Vd\Vd'}{\Dt} - \eta \Big) \dd\beta \right] \wedge \dd\vol(S^2) \,, \\
     \mathcal{D}\beta &= \dd\beta+\frac{2\Vd\Vd'}{2\Vd-\Vdd}\dd\chi, \qquad \Dt = (2\Vd-\Vdd)V''+(\Vd')^2 \,. 
\end{aligned}
\end{equation}
For later use we have written the three-form potential for the four-form field strength rather than the field strength itself, the field strength is of course just the exterior derivative of the three-form given above.

We emphasize that in performing a Backl\"und transform, we are imposing some additional symmetry beyond what is required by supersymmetry and the equations of motion. A priori, it looks unnatural to consider the solutions with this additional $U(1)$ as the Argyres--Douglas theories do not have such a $U(1)$ flavor symmetry. However, as has been discussed in \cite{Bah:2021hei}, this putative $U(1)$ flavor symmetry is broken due to a novel Stuckelberg mechanism when computing anomaly inflow. In \cite{Bomans:2023ouw} this ambiguity was approached from a different viewpoint. There supergravity solutions where this additional $U(1)$ is explicitly broken in the solution were constructed, that is bona fide solutions of the Toda equation rather than the Laplace equation were found. One therefore does not need to rely on the Stuckelberg mechanism to match the symmetries in this case, there is no such isometry to begin with! Despite having a less symmetric solution, \cite{Bomans:2023ouw} showed that the observables computed from both solutions agree exactly and therefore one can reliably work with the Backl\"und transformed solutions.

The utility of the Backl\"und transformation is that the non-linear Toda equation is replaced by the linear, and simpler to solve, 3d cylindrical Laplace equation. Given a solution $V$ to the 3d cylindrical Laplace equation one must impose boundary conditions to obtain a compact solution. Due to the $\rho^2$ term appearing in the metric we may bound the $\rho$ coordinate from below by $0$, where the metric degenerates smoothly.\footnote{It may be possible to find solutions which avoid having $\rho=0$ as a boundary condition altogether, however for the solutions we consider, and all known in the literature, this boundary condition is imposed.} With this boundary condition in tow it is convenient to define the \emph{line charge}:
\begin{equation}
    \lambda(\eta)\equiv \dot{V}(\eta, \rho=0)\,.
\end{equation}
The is a piecewise linear function, and in general takes the form:
\begin{equation}
    \lambda_i(\eta)=\widehat{\alpha}_i \eta+\widehat{\beta}_i\, ,\quad i-1 \leq \eta\leq i\, ,
\end{equation}
with $\widehat{\alpha}, \widehat{\beta}$ constants. 
Following the analysis in \cite{Couzens:2022yjl}, which extends the original analysis in \cite{Gaiotto:2009gz} to allow for the $\D_p(A)$ theories, the line charge must satisfy a number of conditions:
\begin{itemize}
    \item The line charge is a continuous, convex function.
    \item Kinks in the line charge appear only at integer values of $\eta$. 
    \item The constant piece of the line charge, denoted $\widehat{\beta}_i$ above is integer.
\end{itemize}
The origin of these rules will be reviewed later when we consider how they need to be modified for the $BCD$ theories. It is important to emphasize that the slopes of the line charge need not be integer, only the differences in the slopes are quantized. This more restrictive integral condition for the slopes is imposed in \cite{Gaiotto:2009gz}, and when it is imposed it allows for a linear quiver to be constructed. In order to study the $\D_p(G)$ theories, which have no quiver description generically, this must be relaxed and one is forced to consider rational slopes.  

\subsection{Review of the \texorpdfstring{$\D_p(A)$}{DpA} theory}\label{sec:DpAreview}

To get a better feel for the class of solutions, we will review the holographic duals of the Higgsed $\D_p(A)$ theories as discussed in \cite{Couzens:2022yjl,Bah:2022yjf}.  One begins by wrapping $N$ M5-branes on a punctured $\mathbb{CP}^1$, where one puncture is of irregular type and the other puncture, at the antipodal point, is of regular type. The regular puncture is realized by introducing M5-branes intersecting the sphere at a point and is the type of puncture which gives rise to the familiar punctures of higher genus Riemann surfaces in holographic class $\mathcal{S}$ constructions. The irregular puncture is realized differently, arising due to a smeared M5-brane stack which wraps the shrinking circle of the $\mathbb{CP}^1$ at the other pole, \cite{Bah:2021mzw,Bah:2021hei,Couzens:2022yjl,Bah:2022yjf,Bomans:2023ouw}.

The UV the theory is simply a stack of $N$ M5-branes located at the tip of $\mathbb{R}^5$, with the $SO(5)$ isometry of $\mathbb{R}^5$ realized by the $S^4$ link of the cone. This is of course the R-symmetry of the 6d $\mathcal{N}=(2,0)$ theory of type A. Upon wrapping the M5-branes on the punctured $\mathbb{CP}^1$, the geometry is topologically $\mathbb{R}^{1,3}\times \mathbb{CP}^1\ltimes \mathbb{R}^5$, where the $\mathbb{CP}^1$ is fibered over the $S^4$ encoding the necessary twisting of the 6d theory. Flowing to the IR we obtain the near-horizon geometry 
\begin{equation}
    \text{AdS}_5\times \mathbb{CP}^1 \ltimes \widetilde{S}^4 \,,
\end{equation}
where, due to the fibration and warping, the metric on $\widetilde{S}^4$ no longer has the full $SO(5)$ isometry, rather only an $SU(2)$ subgroup remains. We will refer to the compactification $\mathbb{CP}^{1}$ as the punctured two sphere and disc interchangeably. 

This of course agrees with the expectation of a 4d $\mathcal{N}=2$ SCFT living on the compactified $N$ M5-branes, and is the solution studied in \cite{Bah:2021hei,Bah:2021mzw,Couzens:2022yjl, Bah:2022yjf,Bomans:2023ouw} whose explicit form we review below. There are $N$ units of four-form flux threading through the $S^4$ and the conserved charges threading through the other four-cycles in the geometry encode the regular and irregular puncture data. It has been shown that these solutions reproduce the leading order anomalies of the Higgsed $\D_p(A)$ theories. 

To give the solution it suffices to give the potential $V$. 
It is made up of a set of building blocks, which we call $\mathcal{V}_a$:
\begin{equation}
    \begin{split}
        \mathcal{V}_1(\eta,\rho)&=-\eta \log \rho\, ,\\
        \mathcal{V}_2(\eta,\rho: n)&=\frac{\eta}{\widetilde{w}(\eta,\rho:n)}-\frac{\eta}{2}\log\frac{\widetilde{w}(\eta,\rho:n)+1}{\widetilde{w}(\eta,\rho:n)-1}\, ,\\
        \mathcal{V}_3(\eta,\rho:n)&=-\frac{n}{2}\log\frac{ n \widetilde{w}(\eta,\rho:n)+\eta}{n \widetilde{w}(\eta,\rho:n)-\eta} \,, 
    \end{split}
\end{equation}
where we have defined the function $\widetilde{w}(\eta, \rho:n)$ to be:
\begin{equation}
   \widetilde{w}(\eta,\rho:n)= \frac{1}{\sqrt{2} n}\sqrt{n^2+\rho^2+\eta^2+\sqrt{4 n^2 \rho^2+(\rho^2+\eta^2-n^2)^2}}\, .
\end{equation}
It is not hard to show that the line charges from these three building blocks are:
\begin{equation}
    \begin{split}
        \lambda_1(\eta)&=-\eta\, ,\\
        \lambda_2(\eta)&=
        \begin{cases}
           \eta&0\leq\eta\leq n\, , \\
           0 &0<n\leq \eta\, ,
           \end{cases}\\
           \lambda_3(\eta)&=
        \begin{cases}
           0&0\leq\eta\leq n\, , \\
           n &0<n\leq \eta\, .
           \end{cases}
    \end{split}
\end{equation}
In order to describe the Higgsed $\D_p(A)$ theories one needs to take suitable linear combinations of the above building blocks. It is convenient to define $\mathcal{V}_{\text{reg}}(\eta,\rho: n)=\mathcal{V}_2(\eta,\rho: n)+\mathcal{V}_3(\eta,\rho: n)$ which gives rise to the piecewise continuous line charge
\begin{equation}
    \lambda_{\text{reg}}^{\text{block}}(\eta)=\begin{cases}
        \eta &0\leq \eta\leq n\, ,\\
        n &n\leq \eta\, .
    \end{cases}
\end{equation}
By construction this satisfies the necessary regularity conditions given above, including after multiplication by an integer. Expanding the metric around the kink of the line charge at $\eta=n$ gives an internal space of the form $S^2\times \mathbb{R}^4/\mathbb{Z}_{k}$, where $k$ is the change of the slope at the kink (for the above example this is $1$). As in the standard holographic dictionary, this leads to an $SU(k)$ flavor symmetry in the dual field theory. We will revisit how to realize this later in Section \ref{sec:anomalyinflow}.

To describe an arbitrary partition $[i^{m_i}]$, of $N=\sum_{i=1}^{N} i m_i$, we take a judicious superposition of the above regular building blocks:
\begin{equation}
    V_{\text{reg},[i^{m_i}]}=\sum_{i=1}^{N} m_i \mathcal{V}_{\text{reg}}(\eta,\rho: i)\, ,
\end{equation}
where the $m_i$ are forced to be positive integers by the regularity conditions.
This gives rise to the line charge:
\begin{equation}
    \lambda_{\text{reg}}(\eta)=\eta \sum_{j=i}^{N}m_j+\sum_{j=1}^{i-1} i m_j\, ,\quad i-1\leq \eta \leq i\,,
\end{equation}
which fully encodes the data of the regular puncture. 

It remains to introduce the irregular puncture data, which is encoded by the first potential $\mathcal{V}_1$. We have
\begin{equation}
    V_{\text{irreg}}(\eta,\rho:p)=- \frac{N}{p}\eta \log \rho\, ,\quad\Rightarrow \quad \lambda_{\text{irreg}}(\eta)=-\frac{N}{p}\eta\, ,
\end{equation}
with $p$ a positive integer
The total potential is the sum of these two terms, 
\begin{equation}
    V_{\text{tot}}(\eta,\rho:p, [i^{m_i}])= V_{\text{irreg}}(\eta,\rho:p)+ V_{\text{reg}}(\eta,\rho:[i^{m_i}])\, ,
\end{equation}
and has line charge:
\begin{equation}\label{eq:SUlinecharge}
    \lambda(\eta)=\eta\left(\sum_{j=i}^{N}m_j-\frac{N}{p}\right)+\sum_{j=1}^{i-1}j m_j\, ,\quad i-1\leq \eta\leq i\, .
\end{equation}
A few comments are in order. Firstly, by construction, the line charge is a piecewise continuous function with kinks at integer values of $\eta$. Moreover, it has two zeroes, one at $\eta=0$ and the second at $\eta=p$, with $p$ integer. It is interesting to note that there is no requirement for $p$ to be greater than $N$, nor, as a matter of fact, any of the $i$ less than $N$ for which $m_i\neq0$. The solution is consistent regardless provided we impose $p>1$ and that it is integer. When $p<N$ we may interpret this as an obstruction to being able to fully Higgs the full puncture. If $p<N$ then one is restricted to a Higgsing that satisfies
\begin{equation}
    N =\sum_{j=1}^{N}  m_j \min(j,p) \, .
\end{equation}
We are not aware of a similar result in the field theory literature and it would be interesting to derive this directly there. 

The final comment we need to make is to explain the boundary conditions of the full solution. We have seen that at $\rho=0$ we have a shrinking circle and our line charge located there. Moreover along $\rho=0$ the $\eta$ coordinate is restricted to lie between $[0,p]$. We must now fix the range in the full $\eta,\rho$-plane. Computing $\dot{V}_{\text{tot}}$ one finds 
\begin{equation}
    \dot{V}_{\text{tot}}(\eta,\rho: p,[i^{m_i}])=-\frac{N}{p}\eta+\eta\sum_{i=1}^{N}\frac{m_i}{\widetilde{w}(\eta,\rho:i)}\, .
\end{equation}
It follows simply that $\dot{V}|_{\eta=0}=0$, and with a little analysis of the metric one finds that this is a smooth degeneration of the metric. The final boundary condition is imposed by setting $\dot{V}=0$ along some contour, with $\Gamma$ in the positive quadrant. This contour $\Gamma$ satisfies:
\begin{equation}
    0=-\frac{N}{p}+\sum_{i=1}^{N}\frac{m_i}{\widetilde{w}(\eta,\rho:i)}\, .
\end{equation}
For a single non-zero $m_i$ this is the equation of an ellipse:
\begin{equation}
    (m_i^2 p^2-N^2)\eta^2+m_i^2 p^2=\frac{m_i^2 i^2p^2(m_i^2 p^2-N^2)}{N^2}\, .
\end{equation}
For a more generic solution with multiple non-trivial $m_i$ we cannot give a closed form expression for the curve, it is the intersection of the level sets of ellipses, see \cite{Couzens:2022yjl}.

The vanishing of $\dot{V}$ along the contour $\Gamma$ induces a degeneration of the metric which is singular. Despite this, the singularity is physical, corresponding to the locus of a smeared stack of M5-branes. This stack of smeared M5-branes is the holographic realization of the irregular puncture. The smearing here is expected to be a remnant of the supergravity description and lifted by considering higher derivative corrections, resulting in a localized distribution of M5-branes realizing the puncture. Localizing the M5-branes along the smearing circle breaks the electrostatic $U(1)$, as discussed in \cite{Bomans:2023ouw}, however as we mentioned earlier the electrostatic description still has its merits. 

\subsection{\texorpdfstring{$\mathbb{Z}_2$}{Z2} orbifolds in M-theory}\label{sec:Z2}

Having reviewed the holographic dual of the $\D_{p}(A)$ theory we turn our attention to the duals of the $\mathcal{D}_p(BCD)$ theories. Since the most general local $\mathcal{N}=2$ preserving AdS$_5$ solutions are of LLM type the most we can do is to take orbifolds of that setup. The origin of the $\D_p(SO(2N))$ theory, as a compactification of the 6d $\mathcal{N}=(2,0)$ theory on the punctured $\mathbb{CP}^1$ without a twist line, is the most illuminating to consider first. 

The 6d $\mathcal{N}=(2,0)$ $D_N$ theory is realized by placing $2N$ coincident M5-branes on $\mathbb{R}^{1,5}\times\mathbb{R}^5$ at the origin of $\mathbb{R}^5$ and orbifolding the $\mathbb{R}^5$ by the $\mathbb{Z}_2$ parity:
\begin{equation}
    x_i\rightarrow -x_i\, ,
\end{equation}
with $x_i$ the coordinates on $\mathbb{R}^5$. In addition to the action on the coordinates the $\mathbb{Z}_2$ must also act on the antisymmetric three-form potential $C$ as 
\begin{equation}
    C\rightarrow -C\, .
\end{equation}
The orbifold preserves the 6d $\mathcal{N}=(2,0)$ supersymmetry for M5-branes whose world-volume is transverse to $\mathbb{R}^5/\mathbb{Z}_2$ and the fixed plane carries $-1$ unit of M5-brane charge. The net M5-brane charge, as computed in the covering space, is $2N-1$. Flowing to the near-horizon of the M5-branes we obtain the metric $\text{AdS}_7\times\mathbb{RP}^4$. 

The $\D_p(SO(2N))$ theory is obtained by by compactifying the 6d $\mathcal{N}=(2,0)$ $D_N$ above on a punctured $\mathbb{CP}^1$. The pertinent observation is that the theory has the same local form as the holographic duals of the $\D_p(SU(N))$ theory, but there is now a global $\mathbb{Z}_2$ orbifold action on the topological $S^4$. One may see this more clearly by recalling the 7d maximal gauged supergravity origin of the $\D_p(SU(N))$ solutions with rectangular punctures. In this picture the 7d geometry is AdS$_5\times \mathbb{D}_2$ and is uplifted to 11d supergravity on a round $S^4$. The $\D_p(SO(2N))$ is then the same 7d solution uplifted on $\mathbb{RP}^4$ rather than the $S^4$. This is of course the simplest of the other $\D_p(G)$ theories since it does not involve twist lines however it shows that the missing ingredient is a $\mathbb{Z}_2$ quotient of the local solutions.

Given that $\mathbb{Z}_2$ are key to obtaining the other $\D_{p}(G)$ theories, we will review the different $\mathbb{Z}_2$ orbifolds consistent with the symmetries of the problem. There are four such distinct orbifolds, in agreement with the number of distinct $\mathcal{D}_p(G)$ theories we wish to identify the holographic duals of. As we will review, there is a simple connection between the different orbifolds we wish to consider here and the four different types of $O4$ orientifold planes in Type IIA \cite{Hori:1998iv}. There is a natural interpretation of our results from a Type IIA perspective which will be studied elsewhere \cite{CKLLquiver}.

Consider the spacetime geometry $X=\mathbb{R}^{1,4}\times S^1\times \mathbb{R}^5$, where the $S^1$ is understood to be the circle of the disc and one of the coordinates of $\mathbb{R}^{1,4}$ becomes the line interval over which the disc circle is fibered. We want to study the possible $\mathbb{Z}_2$ quotients of this space which also preserve the IR symmetries we require. As such we may only quotient combinations of the $\mathbb{R}^5$ or the $S^1$.

\paragraph{\uline{$\mathbb{R}^5/\mathbb{Z}_2$ quotients:}}

To begin, consider the $\mathbb{Z}_2$ quotient $\gamma$ acting only on $\mathbb{R}^5$. Let $\widetilde{X}=X/\mathbb{Z}_2$. Since the $\mathbb{Z}_2$ action flips the orientation of $X$ it follows that the three-form potential $C_3$ necessarily flips sign under the $\mathbb{Z}_2$ action and therefore so too does its field strength $G_4$. This implies that $G_4$ does not define a four-form but rather a four-form with values in the orientation bundle of $\widetilde{X}$. We may still define the integral of $G$ through a four-cycle of $\widetilde{X}$ by working in the double cover. Let $S$ be a $\mathbb{Z}_2$ invariant submanifold of $X$ with an orientation that is flipped under the $\mathbb{Z}_2$ action. We define the integral of $G$ through its $\mathbb{Z}_2$ quotient $\widehat{S}$ as half the flux through $S$:
\begin{equation}
    \int_{\widehat{S}} G=\frac{1}{2}\int_S G\, .
\end{equation}
Due to the presence of a non-trivial Stiefel--Whitney class the Dirac quantization condition is modified to:
\begin{equation}\label{eq:Dirac}
    2\int_{\widehat{S}} \frac{G}{(2\pi\lp)^3}=\int_{\widehat{S}}w_4(X)\mod 2\, ,
\end{equation}
with $w_4$ the fourth Stiefel--Whitney class of the spacetime. 

Consider a pair of M5-branes in $X$ which are parallel and separated from the $\mathbb{Z}_2$-fixed plane and are the mirrors of each other. It is possible to pick a four-sphere surrounding one of the M5-branes together with its mirror image, $\widehat{S}_{(a)}^{4}$.  Then the cycle $S=S^4_{(1)}\cup S^4_{(2)}$ has its orientation flipped by $\gamma$. The flux through the quotient $\widehat{S}=S/\mathbb{Z}_2$ is then
\begin{equation}
   \int_{\widehat{S}} \frac{1}{2\pi\lp^3}G=\frac{1}{2}\left(\int_{S_{(1)}^{4}} \frac{1}{2\pi\lp^3} G+\int_{S_{(2)}^{4}} \frac{1}{2\pi\lp^3} G\right)=\frac{1+1}{2}=1\, .
\end{equation}
One the other hand the tangent bundle of the spacetime is topologically trivial and therefore $w_4$ vanishes, thereby satisfying the quantization condition. This is precisely the $\mathbb{Z}_2$ action considered for the 6d $D_N$ theory and leads to an $SO(2N)$ gauge theory when there are $2N$ M5-branes. 

Consider now the possibility of including M5-branes on top of the singularity. Since the supergravity approximation can break down here one should remove a neighborhood around the origin of $\mathbb{R}^5/\mathbb{Z}_2$: let this space be $\widetilde{\mathbb{R}^5}/\mathbb{Z}_2$. We can take the $S^4$ surrounding the origin of $\mathbb{R}^5$ which is $\mathbb{Z}_2$ invariant. Then the integral of the flux over the quotient $S=S^4/\mathbb{Z}_2$ is defined as $\tfrac{1}{2}$ of the flux through the $S^4$. The fourth Stielfel--Whitney class does not vanish ($\mod 2$) in this case and the resultant flux quantization condition is:
\begin{equation}
    2\int_S\frac{1}{(2\pi\lp)^3 } G=1\mod 2\,.
\end{equation}
It follows that the total quantized flux through a four sphere surrounding the fixed plane is odd when measured in the double cover. The fixed plane itself carries $-1$ units of M5-brane charge \cite{Dasgupta:1995zm,Witten:1995em,Hori:1998iv} and therefore one can only affix an even number of M5-branes on top of the fixed plane. This gives two distinct configurations to consider. Either a $\mathbb{R}^5/\mathbb{Z}_2$ orbifold or a $\mathbb{R}^5/\mathbb{Z}_2$ orbifold with two M5-branes fixed on top of the fixed plane. The former gives rise to $SO(\text{even})$ gauge groups, and upon reduction to Type IIA the $O4^{-}$ orientifold, while the latter gives rise to $USp(\text{even})$ gauge groups and upon reduction to Type IIA the $O4^{+}$ orientifold. 

\paragraph{\uline{T-rule:}}
So far everything we have discussed will provide gauge groups in 4d rather than flavors. Flavor branes are obtained by intersecting transversely M5-branes along the fixed plane. Concretely these transverse M5-branes share four directions with the fixed plane, $\mathbb{R}^{1,3}\subset \mathbb{R}^{1,4}$ and are transverse to the $S^1$. The t-rule states that only an even number of M5-branes may intersect  the fixed plane transversely, an odd number is inconsistent with flux quantization. 
However, if the fixed plane is screened by a pair of M5-branes lying on top of it then there is no requirement for an even number of transverse M5-branes, an odd number is also consistent. 

\paragraph{\uline{$(\mathbb{R}^5\times S^1)/\mathbb{Z}_2$ quotients:}}
We have now exhausted one type of quotient and we now consider the second. Let the coordinates on $\mathbb{R}^5$ be $x$ and the coordinate on $S^1$ be $\beta$ (this has been chosen for later comparison with the metric in equation \eqref{eq:metric in terms of V in 11d}). Then the $\mathbb{Z}_2$ quotient acts on $\mathbb{R}^5\times S^1$ as:
\begin{equation}
    x\rightarrow -x \, ,\qquad \beta\rightarrow \beta+\pi\, .
\end{equation}
This $\mathbb{Z}_2$ action is free and therefore there is no five-brane charge associated to a fixed point as in the previous case. Upon reducing to Type IIA on the $S^1$ this gives rise to the $\widetilde{O4}^{-}$ orientifold.

We may also wrap an M5-brane on the invariant cylinder $\mathbb{R}^{1,4}\times S^1$ at the center of $\mathbb{R}^5$. Its motion is frozen on the cylinder. One can construct an invariant four-cycle $S=S_0^4\cup S_\pi^4$ in the double cover by taking the four-sphere surrounding the origin of $\mathbb{R}^5$ at $\beta=0$ and $\beta=\pi$ respectively. The flux through $S$ is then $2$, and therefore through the quotient it is $1$. On the other hand the fourth Stiefel--Whitney class vanishes since the $(\mathbb{R}^5\times S^1)/\mathbb{Z}_2$ is homotopy equivalent  to $S^1$ where $w_4=0$ trivially. Upon reduction to Type IIA this gives rise to the $\widetilde{O4}^+$ orientifold.

We have now considered four different types of orbifold, but we want to understand how we can distinguish them apart. 
If we reduce along the circle to Type IIA, we obtain an RR 1-form $A_{\text{RR}}$ which is the connection of the $U(1)$ bundle over the 10d Type IIA spacetime $\mathbb{R}^{1,4}\times \mathbb{R}^5/\mathbb{Z}_2$. Moreover the reduction of the four-form $G$ along the M-theory circle gives rise to a three-form field strength for NS5-branes. We may distinguish between the different classes of orbifold by considering:
\begin{equation}
    \theta \equiv \frac{1}{2\pi}\int_{\mathbb{RP}^2}B\mod 2\, ,\quad \varphi\equiv\frac{1}{2\pi} \int_{S^1}C_1\mod 2\, ,
\end{equation}
with $\mathbb{RP}^2$ and the $S^1$ cycles in $\mathbb{RP}^4$, and $B$ the field strength of the three-form obtained from the reduction. The reduction of the orbifolds where the $S^1$ is undergoes a quotient leads to a RR 1-form with non-trivial holonomy, giving $\varphi=1$. For the orbifolds where the $S^1$ was untouched we have $\varphi=0$. Similarly for the theories where there are no trapped M5-branes we have $\theta=0$, while for those orbifolds including trapped branes we have $\theta=1$. This is summarized in Table \ref{tab:O4s}.

\begin{table}[H]
    \centering
    \begin{tabular}{|c|c|c|c|c|}
    \hline
  M-theory realization &  $O4$ & $(\theta,\varphi)$&Gauge group & Flavor \\
    \hline\hline
  $\mathbb{R}^5/\mathbb{Z}_2\times S^1$ &  $O4^-$    & $(0,0)$ & $SO(\text{even})$ & $USp$\\
     \hline
     $\mathbb{R}^5/\mathbb{Z}_2\times S^1$ and pair of fixed M5s & $O4^+$     &$(1,0)$ &$USp(\text{even})$ & SO\\
    \hline
     $(\mathbb{R}^5\times S^1)/\mathbb{Z}_2$ & $\widetilde{O4}^{-}$ &$(0,1)$&$SO(\text{odd})$& $USp$\\
    \hline
      $(\mathbb{R}^5\times S^1)/\mathbb{Z}_2$ and one trapped M5 & $\widetilde{O4}^{+}$ &$(1,1)$&$USp^\prime(\text{even})$ & SO
\\
\hline
    \end{tabular}
    \caption{The different types of orientifold 4-brane, distinguished by their values of $(\theta,\varphi)$, the gauge group and flavor groups, and their M-theory origin. }
    \label{tab:O4s}
\end{table}

There are two final rules that we must impose for consistency. These are easier to impose in Type IIA first and then consider the uplift to 11d. Firstly, since $B$ is a source for NS5-branes, if an $O4$-plane passes through an NS5-brane its $\theta$ value is shifted by $1$ unit mod $2$. Similarly, since D6-branes are a source for the 1-form RR potential, if an $O4$-plane passes through a D6-brane the value of $\varphi$ is shifted by $1$ mod $2$. These rules agree with the rules provided in \cite{Nishinaka:2012vi} for the type IIA orthosymplectic quiver theories. 

\subsection{Holographic realization of the \texorpdfstring{$\D_p(G)$}{Dp(G)} theories}

Having reviewed the AdS$_5$ solutions dual to the (Higgsed) $\D_p(A)$ theories and the various $\mathbb{Z}_2$ quotients, we turn our attention to realizing the other $\D_p(G)$ theories. To proceed it is instructive to rephrase Table \ref{tbl:irregs} in terms of the asymptotic geometry upon which we place the M5-branes. We have already done this for the $\D_p(A)$ theories and the $\D_p(SO(\text{even}))$ theories in the previous sections. The important ingredient that remains is to realize the twist lines. Fortunately this is quite natural from our discussion of the $\mathbb{Z}_2$ quotients above. 

The $\mathcal{D}_p(USp(2N))$ theory is obtained by performing a $\mathbb{Z}_2$ quotient on $\mathbb{R}^5$ and introducing a pair of fixed M5-branes screening the fixed plane. The total number of M5-branes (in the double cover) is $2N+2$, with two units of M5-brane charge arising from the screening M5-branes. 
In addition there is $-1$ unit of M5-brane due to the fixed plane. 
 
The two remaining theories are twisted compactifications of the 6d $A_N$ theory. The outer-automorphism twist is realized by quotienting the $U(1)$ of the $\mathbb{P}^1$. The $\D_p(SO(2N+1))$ theory contains a $(\mathbb{R}^5\times S^1)/\mathbb{Z}_2$ while the $\D_p(USp'(2N))$ theory contains a $(\mathbb{R}^5\times S^1)/\mathbb{Z}_2$ quotient with one trapped M5-brane. 

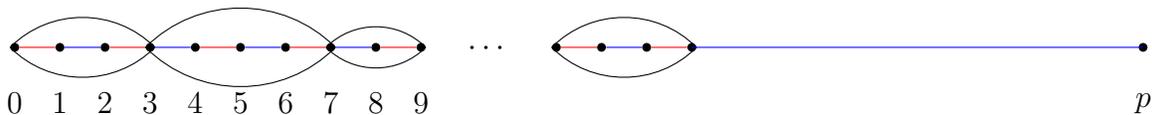
\begin{figure}
    \centering
    \begin{tikzpicture}[scale=0.6]
        \node[draw=none,label={[label distance=-4mm]north:$\cdots$}] (DDD) at (10.5,0) {};
        \node[circle,thick,scale=0.3,fill=black,label={[label distance=4mm]south:$0$}] (A0) at (0,0) {};
        \node[circle,thick,scale=0.3,fill=black,label={[label distance=4mm]south:$1$}] (A1) at (1,0) {};
        \node[circle,thick,scale=0.3,fill=black,label={[label distance=4mm]south:$2$}] (A2) at (2,0) {};
        \node[circle,thick,scale=0.3,fill=black,label={[label distance=4mm]south:$3$}] (A3) at (3,0) {};
        \node[circle,thick,scale=0.3,fill=black,label={[label distance=4mm]south:$4$}] (A4) at (4,0) {};
        \node[circle,thick,scale=0.3,fill=black,label={[label distance=4mm]south:$5$}] (A5) at (5,0) {};
        \node[circle,thick,scale=0.3,fill=black,label={[label distance=4mm]south:$6$}] (A6) at (6,0) {};
        \node[circle,thick,scale=0.3,fill=black,label={[label distance=4mm]south:$7$}] (A7) at (7,0) {};
        \node[circle,thick,scale=0.3,fill=black,label={[label distance=4mm]south:$8$}] (A8) at (8,0) {};
        \node[circle,thick,scale=0.3,fill=black,label={[label distance=4mm]south:$9$}] (A9) at (9,0) {};
        \node[circle,thick,scale=0.3,fill=black] (B1) at (12,0) {};
        \node[circle,thick,scale=0.3,fill=black] (B2) at (13,0) {};
        \node[circle,thick,scale=0.3,fill=black] (B3) at (14,0) {};
        \node[circle,thick,scale=0.3,fill=black] (B4) at (15,0) {};
        \node[circle,thick,scale=0.3,fill=black,label={[label distance=4mm]south:$p$}] (C) at (25,0) {};
        \draw (A0.west) to[bend left=45] (A3.east);
        \draw (A0.west) to[bend right=45] (A3.east);
        \draw (A3.west) to[bend left=45] (A7.east);
        \draw (A3.west) to[bend right=45] (A7.east);
        \draw (A7.west) to[bend left=45] (A9.east);
        \draw (A7.west) to[bend right=45] (A9.east);
        \draw (B1.west) to[bend left=45] (B4.east);
        \draw (B1.west) to[bend right=45] (B4.east);
        \draw[red] (A0)--(A1);
        \draw[blue] (A1)--(A2);
        \draw[red] (A2)--(A3);
        \draw[blue] (A3)--(A4);
        \draw[red] (A4)--(A5);
        \draw[blue] (A5)--(A6);
        \draw[red] (A6)--(A7);
        \draw[blue] (A7)--(A8);
        \draw[red] (A8)--(A9);
        \draw[red] (B1)--(B2);
        \draw[blue] (B2)--(B3);
        \draw[red] (B3)--(B4);
        \draw[blue] (15 ,0)--(25,0);
    \end{tikzpicture}
    \caption{The behavior of the $\mathbb{Z}_2$ quotient as one moves between $\eta = 0$ and $\eta = p$. The red/blue line segments correspond to the different values of $\theta$.}
    \label{fig:figfig}
\end{figure}

The picture one should have is given in Figure \ref{fig:figfig}. We start off at the irregular puncture point at $\eta=p$, which for the sake of clarity is the right-most point on the $\eta$ axis. Between $p$ and the first kink one can construct a four cycle which is the $S^4$ appearing in the UV AdS$_7\times S^4$ solution. This must have the correct $\mathbb{Z}_2$ quotient applied to it, along with the $\beta$ circle if necessary. As one moves to the left the quotient changes according to the rules laid out at the end of Section \ref{sec:Z2}. Probe M5-branes are located at kinks which induces a change in $\varphi$, while at all integer values of $\eta$ less than the first kink the value of $\theta$ changes. 
The t-rule is then the origin of the partition conditions discussed in Sections \ref{sec:SO} and \ref{sec:USp}. One must therefore give the partition data $[i^{m_i}]$ and the values of $(\theta,\varphi)$ in the rightmost segment to fully determine the solution. 

We conjecture that the line charge for each of the theories, with a partition of $N$ flavor symmetry, takes the universal form  
\begin{equation}
\lambda(\eta)=\widetilde{\lambda}-\frac{b}{s_\mathfrak{g} m p}\eta+\eta\sum_{j=i}^{N} m_{j}+\sum_{j=1}^{i-1}j m_j\, ,\quad i-1\leq \eta \leq i\,.
\end{equation}
Here we have introduced the constant $s_\mathfrak{g}$ which is $1$ for SU and SO flavor symmetries and $\tfrac{1}{2}$ for USp flavor symmetry. The parameters $b$ and $m$ are as in Section \ref{sec:Fieldtheory} and $\widetilde{\lambda}=\pm 2$, with $+$ for USp theories and $-$ for SO theories. These are the values that guarantee that the line charge vanishes at $\eta=p$.

One can now proceed to analyse the constraints from flux quantization. One finds that the analysis is virtually identical to the analysis in the $SU$ case and so we give just the final result and refer the reader to \cite{Couzens:2022yjl} for further details. One finds that the location of the kinks must be integer, $p$ is also integer, and the difference between the slopes to be integer. 

\subsection{Anomaly inflow}\label{sec:anomalyinflow}

Having given the solutions, we now want to understand how to compute observables which we can match with the field theory results. We will focus on three observables: $(2a-c)$, $(c-a)$ and the levels of the flavor symmetries in the following, including subleading contributions. Each of these observables can be extracted from a six-form anomaly polynomial in field theory, see equation \eqref{eq:AMIR}. Our goal in this section is to derive the form of the anomaly polynomial by using inflow on the stack of M5-branes. A similar computation has been performed for the class $\mathcal{S}$ theories arising from punctured higher genus Riemann surfaces in \cite{Bah:2019jts}, and for the $\D_p(A)$ theories in \cite{Bah:2021hei}. Our computation must take into account that the manifold on which we are integrating over may have a non-trivial Stiefel--Whitney class, this leads to a modification of the inflow procedure performed in these previous papers.

For the $\D_p(A)$ theory the anomaly inflow utilized the following $12$-form anomaly polynomial:
\begin{equation}
I_{12}=-\frac{1}{6} E_4\wedge E_4 \wedge E_4 -E_4\wedge X_8\, ,
\end{equation}
where $E_4$ is the fully covariantized four-form of the four-form flux $G$, such that
\begin{equation}
    \int_{S^4}E_4=N\,.
\end{equation}

To see why the anomaly inflow procedure requires modification we need only recall the Dirac quantization condition for the four-form field strength, in equation \eqref{eq:Dirac}. Necessarily the four-form $G$ does not define an integral class. The resolution is simple and given in \cite{Witten:1996md}. One should rather covariantize the integral class:
\begin{equation}
  \widehat{E}_4\equiv  \frac{1}{(2\pi\lp)^3}G_4 -\frac{1}{2}w_4=E_4-\frac{1}{2}w_4\,,
\end{equation}
and then perform the anomaly inflow using the $12$-form anomaly polynomial:
\begin{equation}
    \mathcal{I}_{12}=-\frac{1}{6}\widehat{E}_4\wedge \widehat{E}_4\wedge \widehat{E}_4 -\widehat{E}_4 \wedge X_8\,.
\end{equation}
We see that the modification will result in certain shifts compared to the anomaly inflow arguments for the A-type case. One can now integrate over the internal manifold, being careful with the interchanging of the various orbifolds. The result is:
\begin{equation}
    \begin{split}
        48(c-a)&=t_\mathfrak{g}\sum_{i=1}^{N}m_i(\lambda(i)+\alpha_i)\, ,\\
        2a-c&=\frac{t_\mathfrak{g}}{4}\bigg[\sum_{i=1}^{N}\int_{i-1}^{i}(\lambda(\eta)+\alpha_i)^2\dd\eta+\int_{N}^{p}(\lambda(\eta)+\alpha_{N+1})^2\dd\eta\bigg]+2 t_\mathfrak{g}(c-a)\, ,\\
        k_i&= 2 s_\mathfrak{g}(\lambda(i)+\alpha_i)\, ,
    \end{split}
\end{equation}
where $\alpha_i$ is a constant originating from the shifts with $w_4$. We have also introduced $t_\mathfrak{g}$ which is a result of the prescription for computing in the double cover and then dividing by $2$. For $SU$ we have $t_\mathfrak{g}=1$ and $t_\mathfrak{g}=\tfrac{1}{2}$ for both $USp$ and $SO$. 

One can show that with these expressions the observables match with the field theory results from Sections \ref{sec:nilphiggsing} and \ref{sec:Fieldtheory} even to subleading order. To exemplify this we will study punctures associated to rectangular partitions in Section \ref{sec:rect}, but one can use our results to obtain the subleading results for all permissible Higgsings of the theories. 

\section{Example: Higgsing AD SCFTs via rectangular orbits}\label{sec:rect}

To elucidate the matching between the supergravity solutions and their putative dual field theories, we consider some simple families of examples in this section. In particular, we consider all cases where the integer partitions describing the Higgsing have Young tableaux which are rectangular, and we refer to the associated nilpotent orbits as the ``rectangular orbits''. Such nilpotent orbits possess only one parameter, $M$, which describes the length of the Young tableau, and thus the expressions for the conformal anomalies simplify dramatically.

To further increase the simplicity of this section, we focus on the
Argyres--Douglas theories where the irregular puncture does not contribute any
mass parameters or marginal couplings to the SCFT. In particular, we consider the ultraviolet theories 
\begin{equation}
    \mathcal{D}_p^{\, b=h_t}(G \neq USp', O_\text{full}) \,,
\end{equation}
subject to the constraint that $\gcd(p, h_G^\vee) = 1$, which guarantees the absence of mass parameters/marginal couplings \cite{Xie:2019vzr}. 
Under these assumptions, the central charge $a$ simplifies, and we can write the closed form expression:
\begin{equation}\label{eqn:SIMPLEa}
    a^\text{UV} = \frac{(4p-1)(p-1)}{48p}\operatorname{dim}(\mathfrak{g}) \,.
\end{equation}

Before we explicitly compare the central charges between the supergravity duals and the putative Argyres--Douglas duals, we first clarify what we mean by ``leading order'', etc. When considering a theory $\mathcal{D}_p(G)$, with $G = SU(N)$, $SO(N)$, or $USp(N)$, Higgsed by the rectangular partition $[M^{N/M}]$, there are three quantities that appear in the central charges: $N$, $M$, and $p$. The supergravity description on the holographic side provides a good approximation when $N$ and $p$ are both large. Naively, a line charge corresponding to a large value of $M$, e.g., $M \simeq N$, may lead to monomials dominating the central charges that are subleading for $M \simeq 1$. This motivates the following definition of the degree of a monomial:
\begin{equation}\label{eqn:degree}
    \operatorname{deg}\left(p^{n_p} N^{n_N} M^{n_M}\right) = n_p + n_N + \begin{cases}
        n_M &\quad \text{ if } n_M > 0 \\
        0 &\quad \text{ if } n_M < 0 \,.
    \end{cases}
\end{equation}
This definition captures the maximal value for the degree over the range of possible $M$. Often we will use the shorthand $\mathcal{O}(N^d)$, by which we mean monomials that have degree $\leq d$, where the degree is as defined in equation \eqref{eqn:degree}.

\subsection{\texorpdfstring{$\mathfrak{su}(N)$}{su(N)} with \texorpdfstring{$[M^{N/M}]$}{M\textasciicircum N/M}}

In the first instance, we consider the class $\mathcal{S}$ theory with an irregular puncture:
\begin{equation}\label{eqn:eg2}
    \mathcal{D}_p^{\, b = N}\left(SU(N), [M^{N/M}]\right) \,.
\end{equation}
Here, we make the assumption that $N$ is a multiple of $M$, otherwise the partition does not exist; it is also necessary that $p > M$, otherwise the Higgsed theory is typically not an interacting SCFT. The holographic duals of such theories have been studied previously in \cite{Bah:2021mzw,Bah:2021hei}. 

The UV theory from which the theory in equation \eqref{eqn:eg2} is obtained via Higgs branch renormalization group flow is $\mathcal{D}_p^{\, b = N}\left(SU(N), [1^{N}]\right)$, which is also known as $\mathcal{D}_{p}(SU(N))$. We assume that $\gcd(p, N) = 1$; this fixes that the number of mass deformations, $f$, to be zero.\footnote{When $\gcd(p, N) > 1$ then the central charges are the same as below at leading and subleading orders, but not at lower orders. Since we only compare the leading and subleading contributions to the supergravity perspective, so we emphasize once again that the assumption of no mass parameters/marginal couplings is only for the simplicity of the expressions that we write.} Then, the UV theory has conformal anomalies
\begin{equation}
    a^\text{UV} = \frac{1}{48} \frac{(4p-1)(p-1)}{p} (N^2 - 1) \,, \qquad c^\text{UV} = \frac{1}{12} (p - 1)(N^2 - 1) \,.
\end{equation}
Furthermore, the flavor algebra is
\begin{equation}
    \mathfrak{f}_{k^\text{UV}}^\text{UV} = \mathfrak{su}(N)_{2\left(\frac{p - 1}{p}\right) N} \,.
\end{equation}

Now, we determine the required properties of the nilpotent orbit associated to the partition $[M^{N/M}]$. The decomposition is
\begin{equation}
    \mathfrak{su}(N) \rightarrow \mathfrak{su}(2)_X \oplus \mathfrak{su}(m_M = N/M) \,.
\end{equation}
The Dynkin embedding indices are
\begin{equation}
    I_X = \frac{N(M^2 - 1)}{6} \,, \qquad I_M = M \,.
\end{equation}
The coefficients of the Nambu--Goldstone anomaly polynomial are given via the application of equation \eqref{eqn:SUNGHs}. We find
\begin{equation}
    H = \frac{N^2(M-1)}{2M} \,, \quad H_v = \frac{N^2(2M^2(M-2 ) + M + 1)}{6M} \,, \quad H_M = 2N(M-1) \,.
\end{equation}

Now that we have determined the UV anomaly coefficients, and the Dynkin embedding indices and Nambu--Goldstone anomaly coefficients associated to the nilpotent orbit, we are ready to apply equation \eqref{eqn:tachiGB} to find the infrared conformal anomalies. For $(a, c)$, we find
\begin{equation}\label{eqn:SUIRac}
    \begin{aligned}
        12c^\text{IR} &= 12c^\text{UV} - 3I_X k^\text{UV} + 3H_v - H \\
        &= pN^2 + \frac{M^2N^2}{p} - 2MN^2 - p -\frac{N^2}{p} + \frac{N^2}{M} + 1 \,, \\
        48a^\text{IR} &= 48a^\text{UV} - 12I_X k^\text{UV} + 12H_v - 2H \\ 
        &= 4N^2p - 4p + \frac{4M^2N^2}{p} - \frac{3N^2}{p} - \frac{1}{p} - 8MN^2 + \frac{3N^2}{M} + 5 \,.
    \end{aligned}
\end{equation}
For convenience, we also write the difference of the central charges
\begin{equation}\label{eqn:a-crec}
    48(c^\text{IR} - a^\text{IR}) = \frac{N^2}{M} - \frac{N^2}{p} + \frac{1}{p} - 1 \,.
\end{equation}
Furthermore, the flavor symmetry and the flavor central charge of the infrared theory is
\begin{equation}\label{eqn:kfrec}
\begin{aligned}
    \mathfrak{f}^\text{IR}_{k_M^\text{IR}} = &\, \mathfrak{su}(N/M)_{k_M^\text{IR}}\,, \\
    k_M^\text{IR} = &\, I_M k^\text{UV} - H_M = 2N\left(1 - \frac{M}{p}\right) \,.
\end{aligned}
\end{equation}

Now, we wish to demonstrate that these 't Hooft anomalies can be reproduced from the putative dual supergravity solution. For a rectangular puncture $[M^{N/M}]$ the line charge is given by
\begin{equation}
    \lambda(\eta)=\begin{cases}
        \dfrac{N(p_f - M)}{M p_f}\eta & 0\leq \eta\leq M\\[15pt]
        \dfrac{N}{p_f}(p_f - \eta ) & M\leq\eta \leq p_f,
    \end{cases}
\end{equation}
where $p_f$ is the final point of line charge and is associated with the irregular puncture by the holographic dictionary, $p=p_f$. There are two contributions to $(2a - c)$, as discussed in Section \ref{sec:holoDpG}. Both contributions can be computed by using anomaly inflow, the first arises from the canonical Chern--Simons term of 11d supergravity whilst the second arises from the higher derivative corrections involving the $X_8$ anomaly polynomial. We will refer to the former as the $2\partial$ contribution and the latter as the higher derivative correction. Applying the holographic dictionary, the $2\partial$ contribution is 
\begin{equation}\label{eqn:SUR2}
    12 (2a^\text{IR} - c^\text{IR})^\textrm{inflow} \big|_{2\partial} = - 3 \int^{p}_0\lambda(\eta)^2 \dd \eta= - \frac{N^2}{p} (p-M)^2 \,.
\end{equation}
The reproduces the leading order (degree three) terms in the conformal anomalies as determined from the field theory perspective in equation \eqref{eqn:SUIRac}. 

Next, we consider the second contribution which comes from the $C_3 \wedge X_8$ higher derivative Chern--Simons term in M-theory:
\begin{equation}\label{eqn:SUCS}
    12 (2a^\text{IR} - c^\text{IR})^\textrm{inflow} \big|_{X_8} = - \frac{N}{2M} \lambda(M) = - \frac{N^2}{2} \left( \frac{1}{M} - \frac{1}{p} \right) \,.
\end{equation}
Combining both contributions, the $2\partial$ and the $X_8$, we can match the $(2a-c)$ at both leading and subleading order. Comparing the field theory result in equation \eqref{eqn:SUIRac} with the supergravity contributions to the anomaly inflow in equations \eqref{eqn:SUR2} and \eqref{eqn:SUCS}, we find
\begin{equation}
    12(2a^\text{IR}-c^\text{IR}) + 12(2a^\text{IR}-c^\text{IR})^\textrm{inflow} = - p + \frac{3}{2} - \frac{1}{2p} \,.
\end{equation} 
As we can see, the two perspectives begin to diverge at $\mathcal{O}(N)$. This is not unexpected, as the subsubleading terms in the central charges are sensitive to the internal properties of the irregular puncture, such as the additional mass parameters discussed in Section \ref{sec:Fieldtheory}. Moreover, the gravity computation really sees a $U(N)$ flavor symmetry rather than an $SU(N)$ flavor symmetry, similar to the usual story for the gauge group of $\mathcal{N}=4$ SYM. 

We can also compute the contribution to $(c-a)$ from the higher derivative Chern--Simons term. For the rectangular puncture $[M^{N/M}]$ this contribution is
\begin{equation}\label{eqn:SUXXX}
    48(c^\text{IR}-a^\text{IR})^\textrm{inflow} \Big|_{X_8} = - \frac{N}{M} \lambda(M) = - N^2 \left( \frac{1}{M} - \frac{1}{p} \right) \,,
\end{equation}
This is a special case of the general result for the holographic central charges that was obtained from the higher derivative correction in Section \ref{sec:anomalyinflow}. We can compare the difference of central charges to the field theory result in equation \eqref{eqn:a-crec} and find that
\begin{equation}
    48(c-a) + 48(c-a)^\textrm{inflow} \big|_{X_8} = \frac{1}{p} - 1 \,.
\end{equation}
As we can see, this matches at both leading and subleading orders, with the difference arising only at $\mathcal{O}(1)$. One can see that this contribution is actually the contribution for N=1 to $(c-a)^{\text{UV}}$ and arises because gravity sees unitary groups rather than special unitary groups. Subtracting off the contribution of the decoupled $U(1)$ mode gives a perfect match including order one terms. 

Furthermore, we can compare the flavor central charge in equation \eqref{eqn:kfrec} from the field theory perspective with the result from the anomaly inflow method:
\begin{equation}
        (k_M^\text{IR})^\textrm{inflow} = - 2N \left( 1-\frac{M}{p} \right)  \qquad \Rightarrow \qquad k_M^\text{IR} + (k_M^\text{IR})^\textrm{inflow} = 0 \,.
\end{equation}
This is a remarkably precise matching between the field theory and the putative holographic dual. We emphasize that, even in this case which does not incorporate orbifolding, we have demonstrated that the central charges match to higher order than has been shown in the previous literature \cite{Bah:2021mzw,Bah:2021hei,Bah:2022yjf,Couzens:2022yjl}, providing further evidence for the proposed holographic duality. 

\subsection{\texorpdfstring{$\mathfrak{so}(N)$}{so(N)} with \texorpdfstring{$[M^{N/M}]$}{M\textasciicircum N/M}}

Now that we have considered the rectangular partitions for $\mathfrak{su}(N)$,
we now turn to the analogous rectangular partitions for $\mathfrak{so}(N)$.
There are two cases that we should consider here, depending on whether $N$ is
an even or an odd integer. 

The first case we study is where $N$ is even, in which case the class $\mathcal{S}$ theory that we want to compute the conformal anomalies of is
\begin{equation}\label{eqn:eg3}
    \mathcal{D}_p^{\, b = N-2}\left(SO(N), [M^{N/M}]\right) \,.
\end{equation}
It is necessary that $N$ is an integer multiple of $M$ for $[M^{N/M}]$ to be an integer partition of $N$. Furthermore, for the class $\mathcal{S}$ construction to make sense, the partition needs to correspond to a nilpotent orbit of $\mathfrak{so}(N)$; this necessitates that if $M$ is an even integer then $N/M$ is also an even integer. The theory in equation \eqref{eqn:eg3} arises from a Higgs branch renormalization group flow of the class $\mathcal{S}$ theory
\begin{equation}
    \mathcal{D}_p^{\, b = N-2}\left(SO(N), [1^N]\right) \,,
\end{equation}
which is frequently denoted as $\mathcal{D}_{p}(SO(N))$. As in the $\mathfrak{su}(N)$ case, we suppose that 
\begin{align}
    \gcd(p, N-2) = 1, 
\end{align}
which sets the number of mass parameters to be zero: $f=0$. The UV anomalies are then straightforwardly read off from Section \ref{sec:Fieldtheory}.

The second class of theories which we wish to consider in this subsection are
\begin{equation}\label{eqn:eg35}
    \mathcal{D}_p^{\, b = 2N-4}\left(SO(N), [M^{N/M}]\right) \,,
\end{equation}
where $N$ is an odd integer. Since $N$ is odd, then a partition of the form
$[M^{N/M}]$ only exists when $M$ is also odd. The class $\mathcal{S}$ theory
that flows under Higgs branch flow to the SCFT in equation \eqref{eqn:eg35} is
\begin{equation}
    \mathcal{D}_p^{\, b = 2N-4}\left(SO(N), [1^N]\right) \,.
\end{equation}
Again we determine the conformal anomalies of this theory, also known as $\mathcal{D}_p^{2N-4}(SO(N))$, following Section \ref{sec:Fieldtheory}. In fact, in the absence of mass parameters and marginal couplings, the anomalies of the theories $\mathcal{D}_{p}(SO(N))$ and $\mathcal{D}_p^{2N-4}(SO(N))$ take exactly the same form. In particular, we have 
\begin{equation}
\begin{aligned}
     a^\text{UV} & = \frac{1}{48} \frac{(4p-1)(p-1)}{p} \frac{N(N - 1)}{2} \,,\\
     c^\text{UV} & = \frac{1}{12} (p-1)\frac{N(N - 1)}{2} \,.
\end{aligned}
 \end{equation}
Furthermore, the flavor algebra in both cases is
\begin{equation}
     \mathfrak{f}_{k^\text{UV}}^\text{UV} = \mathfrak{so}(N)_{2\left(\frac{p-1}{p}\right) (N-2)} \,.
\end{equation}

We now work out the relevant quantities related to the nilpotent orbit
$[M^{N/M}]$ that we use to trigger the Higgs branch flow. The decomposition is 
\begin{equation}\label{eqn:sobranch}
    \mathfrak{so}(N) \rightarrow \mathfrak{su}(2)_X \oplus \mathfrak{f}_M \quad \text{ where } \quad \mathfrak{f}_M = \begin{cases} 
    \mathfrak{usp}(N/M) &\text{ if $M$ even } \\
    \mathfrak{so}(N/M) &\text{ if $M$ odd. }
    \end{cases}
\end{equation}
The Dynkin embedding indices of these two factors are
\begin{equation}
    I_X = \frac{1}{12} N(M^2 - 1) \,, \qquad I_M = \begin{cases} M/2 \quad &\text{if $M$ even,} \\ M \quad &\text{if $M$ odd.} \end{cases}
\end{equation}
It is straightforward now to use equation \eqref{eqn:adjdecompSO} to write the decomposition of the adjoint representation of $\mathfrak{so}(N)$ under this decomposition. We find that when $M$ is even we have
\begin{equation}\label{eqn:adjMeven}
    \textbf{adj} \rightarrow (\bm{3} \oplus \bm{7} \oplus \cdots \oplus \bm{2M-1}, \bm{1} \oplus \bm{A^2}) \oplus (\bm{1} \oplus \bm{5} \oplus \cdots \oplus \bm{2M-3}, \bm{S^2}) \,.
\end{equation}
Here, $\bm{A^2}$ and $\bm{S^2}$ are, respectively, the second-antisymmetric and the second-symmetric representations of $\mathfrak{usp}(N/M)$. For $M$ odd, we can equally consider the branching rule of the adjoint under the decomposition in equation \eqref{eqn:sobranch}, and we find
\begin{equation}\label{eqn:adjModd}
    \textbf{adj} \rightarrow (\bm{3} \oplus \bm{7} \oplus \cdots \oplus \bm{2M-3}, \bm{1} \oplus \bm{S^2}) \oplus (\bm{1} \oplus \bm{5} \oplus \cdots \oplus \bm{2M-1}, \bm{A^2}) \,,
\end{equation}
where $\bm{A^2}$ and $\bm{S^2}$ are now the second-antisymmetric and the second-symmetric representations of $\mathfrak{so}(N/M)$, respectively.

From the branching rules in equations \eqref{eqn:adjMeven} and \eqref{eqn:adjModd}, it is straightforward to extract the anomaly contributions from the Nambu--Goldstone modes. We write the anomaly coefficients in the following suggestive forms:
\begin{equation}
  \begin{aligned}
    H &= \frac{1}{2} \bigg[ \frac{N^2(M-1)}{2M} \bigg] - \frac{N}{4} + \begin{cases} \frac{N}{4M} \quad &\text{if $M$ odd} \\
    0 \quad &\text{if $M$ even, } \end{cases} \\
    H_v &= \frac{1}{2} \bigg[\frac{N^2(M-1)(2M(M-1) - 1)}{6M}\bigg] + \frac{N(1 + 6M - 4M^2)}{12} - \begin{cases} \frac{N}{4M} \quad &\text{if $M$ odd} \\
    0 \quad &\text{if $M$ even, } \end{cases} \\
    H_M &= \begin{cases} 2N(M-1) - 4(M-1) \quad &\text{if $M$ odd} \\
    N(M-1) - 2M \quad &\text{if $M$ even.} \end{cases}
  \end{aligned}
\end{equation}
The terms in square brackets are, respectively, the $H$ and $H_v$ associated to $\mathfrak{su}(N)$ Higgsed by the partition $[M^{N/M}]$.

Using these quantities, combined as in equation \eqref{eqn:tachiGB}, we find
the conformal anomalies of the SCFTs constructed via the data in equations
\eqref{eqn:eg3} and \eqref{eqn:eg35}.\footnote{We emphasize once again that the
result is the same for both families theories, as the expressions for the UV
anomalies and the Nambu--Goldstone contributions do not depend on the parity of
$N$.} For the central charges $a$ and $c$ in the infrared, we find
\begin{equation}\label{eqn:c,arecSO}
    \begin{aligned}
	    12c^\text{IR} &= \frac{N^2p}{2} - \frac{Np}{2} + \frac{M^2N^2}{2p} - \frac{N^2}{2p} - \frac{M^2N}{p} + \frac{N}{p} \\&\qquad\qquad\qquad - MN^2 + \frac{N^2}{2M} + \frac{3MN}{2} - \begin{cases} \frac{N}{M}  &\quad\text{ if $M$ odd} \\ 0 &\quad\text{ if $M$ even, } \end{cases} \\
        48a^\text{IR} &= 2N^2p - 2Np + \frac{2M^2N^2}{p} - \frac{3N^2}{2p} - \frac{4M^2N}{p} + \frac{7N}{2p} \\&\qquad\qquad\qquad - 4MN^2 + \frac{3N^2}{2M} + 6MN - \begin{cases} \frac{7N}{2M}  &\quad\text{ if $M$ odd} \\ 0 &\quad\text{ if $M$ even. } \end{cases}
    \end{aligned}
\end{equation}
We can also determine the difference of the central charges. We find that it is:
\begin{equation}\label{eqn:cmarecSO}
    48(c^\text{IR} - a^\text{IR}) = \frac{N^2}{2M} + \frac{N}{2p} - \frac{N^2}{2p} - \begin{cases} \frac{N}{2M}  &\quad\text{ if $M$ odd} \\ 0 &\quad\text{ if $M$ even. } \end{cases}
\end{equation}
Finally, when $M$ is even, we note that the infrared theory has a $\mathfrak{usp}(N/M)$ with flavor central charge
\begin{equation}\label{eqn:kfrecSOeven}
    k^\text{IR}_M = N + \frac{2M}{p} - \frac{NM}{p} \,,
\end{equation}
and when $M$ is odd, the infrared theory has an $\mathfrak{so}(N/M)$ flavor algebra with flavor central charge
\begin{equation}\label{eqn:kfrecSOodd}
    k^\text{IR}_M = 2N + \frac{4M}{p} - \frac{2NM}{p} - 4 \,.
\end{equation}

Now, we wish to compare these conformal anomalies to those obtained from the supergravity duals using the holographic dictionary. In Section \ref{sec:holoDpG}, we argued that the piecewise linear line charge when $\mathfrak{g} = \mathfrak{so}(N)$ is given as
\begin{equation}\label{eqn:lineline}
    \lambda(\eta)= -2 + \eta \left( \sum_{j=i}^N m_j - \frac{b}{m p s_G} \right) + \sum_{j=1}^{i-1} j m_j \,,
\end{equation}
for $\eta$ in the range $i - 1 \leq \eta \leq i$. We first determine the two-derivative contribution to $(2a - c)$ from the anomaly inflow discussed in Section \ref{sec:holoDpG}. We find
\begin{equation}
  \begin{aligned}
    12(2a^\text{IR}-c^\text{IR})^\textrm{inflow} \big|_{2\partial} &= - \frac{3}{2} \int_0^p \lambda(\eta)^2 d\eta \\ 
    &= MN^2 - \frac{M^2 N^2}{2p}  - \frac{N^2 p}{2} + \frac{M^2 N}{p} - 3MN + 2Np  - 2p  \,.
  \end{aligned}
\end{equation}
As we can see, the three leading order, i.e., $\mathcal{O}(N^3)$, terms here match with the leading order contributions appearing in equation \eqref{eqn:c,arecSO}. The subleading, $\mathcal{O}(N^2)$, terms do not match, as there are additional contributions to $(2a - c)$ coming from the higher-derivative corrections due to $X_8$.

Now that we have determined the leading order contribution to $(2a - c)$, we want to determine the leading and subleading contributions to the gravitational anomaly $(c-a)$. Recall that, for a Higgsing by an arbitrary B/D-partition of $N$, we found that
\begin{equation}
    48(c^\text{IR}-a^\text{IR})^\text{inflow} = -\frac{1}{2} \sum_{i=1}^N m_i \left( \lambda(i) + \alpha_i \right) \,,
\end{equation}
we can see that when we have a rectangular partition, we have
\begin{equation}
    48(c^\text{IR}-a^\text{IR})^\text{inflow} = -\frac{1}{2} \left[ - \frac{2N}{M} +  N^2 - \frac{N^2 (M - 1)}{M} - \frac{N b}{m p s_G} + \frac{N}{M}\alpha_M  \right] \,.
\end{equation}
Finally, we recall that $\alpha_i = 1$ if $i$ is odd, and $2$ otherwise. Therefore, we find that the supergravity calculation tells us that
\begin{equation}
    48(c^\text{IR}-a^\text{IR}) = \frac{N^2}{2M} - \frac{N^2}{2p} + \frac{N}{p} + \frac{N}{2M}(\alpha_M-2) \,,
\end{equation}
which matches the field theory computation in equation \eqref{eqn:cmarecSO}. We note that this is an matching at $\mathcal{O}(N^2)$ and $\mathcal{O}(N)$, but not at subleading order where we find that the difference is
\begin{equation}
    48(c^\text{IR}-a^\text{IR}) + 48(c^\text{IR}-a^\text{IR})^\text{inflow} = - \frac{N}{2p} \,.
\end{equation}
This is a UV contribution, as it is independent of the choice of nilpotent orbit describing the Higgsing. Recall that we have assumed that the field theory satisfies $\gcd(p, h_\mathfrak{g}^\vee) = 1$; when this condition is relaxed there are extra contributions from the irregular puncture, which affect both the supergravity and the field theory results.

Next, we turn to the matching of the flavor central charge. The general formula, in terms of the line charge, is
\begin{equation}
    k_i^\text{inflow} = -2s_G^{(i)} \lambda(i) \,,
\end{equation}
where $s_G^{(i)}$ is $1$ if the $i$th flavor algebra is of $\mathfrak{so}$-type, and $1/2$ is it is $\mathfrak{usp}$. Applying this to the flavor symmetry after Higgsing by the rectangular partition, we find 
\begin{equation}
    k_M^\text{inflow} = -\begin{cases}
        N + \dfrac{2M}{p} - \dfrac{MN}{p} + 2 &\qquad \text{if $M$ even,} \\[15pt]
        2N + \dfrac{4M}{p} - \dfrac{2MN}{p} + 4 &\qquad \text{if $M$ odd.} \\
    \end{cases}
\end{equation}
Comparing this to the field theory result, we find that the difference is
\begin{equation}
    k_M^\text{IR} + k_M^\text{inflow} = \begin{cases}
        2 &\qquad \text{if $M$ even,} \\
        0 &\qquad \text{if $M$ odd.} \\
    \end{cases}
\end{equation}
Therefore, while we have an $\mathcal{O}(N)$ matching, we do not quite have an $\mathcal{O}(1)$ matching, due to this factor of $2$ when $M$ is even. This subleading order mismatch does not depend on $M$, and therefore it is unrelated to the Higgsing: it is a UV mismatch.\footnote{It is also not special to configurations involving irregular punctures; it also appears for linear orthosymplectic quivers, as discussed in \cite{CKLLquiver}.}

\subsection{\texorpdfstring{$\mathfrak{usp}(N)$}{usp(N)} with \texorpdfstring{$[M^{N/M}]$}{M\textasciicircum N/M}}

Finally, we wish to consider the conformal anomalies for Argyres--Douglas theories where a ultraviolet $\mathfrak{usp}(N)$ flavor symmetry has been Higgsed by giving a nilpotent vacuum expectation value to the moment map operator which is associated to the rectangular C-partition $[M^{N/M}]$.\footnote{It is necessary that $N$ is even and that $N/M$ is an integer when $M$ is even and an even integer when $M$ is odd; the latter condition is necessary to satisfy the C-partition constraint.} We introduced two classes of UV theories with a $\mathfrak{usp}(N)$ flavor symmetry, which were denoted
\begin{equation}\label{eqn:uspRUV}
\mathcal{D}_p^{\, b}\left(USp(N), [M^{N/M}]\right) \quad \text{ and } \quad \mathcal{D}_p^{\, b}\left(USp'(N), [M^{N/M}]\right)
    \,.
\end{equation}
However, we do not consider the $USp'$ theory in this section, as it does not have the simplification described in equation \eqref{eqn:SIMPLEa}.
We then use the UV data, as worked out in Section \ref{sec:Fieldtheory}, together with the nilpotent Higgsing procedure laid-out in Section \ref{sec:nilphiggsing}, to determine the conformal anomalies of the theory 
\begin{equation}\label{eqn:uspRthy1}
\mathcal{D}_p^{\, b=N+2}\left(USp(N), [M^{N/M}]\right)
    \,.
\end{equation}
For ease of reading the UV anomalies are:
\begin{equation}
    a^{\text{UV}}=\frac{1}{48}\frac{(4p-1)(p-1)}{p}\frac{N(N+1)}{2}\, ,\quad c^{\text{UV}}=\frac{1}{12}(p-1)\frac{N(N+1)}{2}\,,
\end{equation}
while the flavor algebra is
\begin{equation}
\mathfrak{f}_{k^{\text{UV}}}^{\text{UV}}=\mathfrak{usp}(N)_{\frac{p-1}{p}(N+2)}\, .
\end{equation}

We first discuss the contributions to the conformal anomalies arising from the Nambu--Goldstone modes. The nilpotent orbit, $[M^{N/M}]$ of $\mathfrak{usp}(N)$, is associated to the decomposition
\begin{equation}\label{eqn:uspRdecomp}
    \mathfrak{usp}(N) \rightarrow \mathfrak{su}(2)_X \oplus \mathfrak{f}_M \quad \text{ where } \quad \mathfrak{f}_M = \begin{cases} 
    \mathfrak{usp}(N/M) &\text{ if $M$ odd } \\
    \mathfrak{so}(N/M) &\text{ if $M$ even, }
    \end{cases}
\end{equation}
where the Dynkin embedding index of the $\mathfrak{su}(2)_X$ factor is
\begin{equation}
    I_X = \frac{N(M^2 - 1)}{6} \,.
\end{equation}
It is then straightforward to read off the branching of the adjoint representation via equation \eqref{eqn:adjdecompUSP}.
When $M$ is even the adjoint representation of $\mathfrak{usp}(N)$ has the following branching rule under the decomposition in equation \eqref{eqn:uspRdecomp}:
\begin{equation}
    \textbf{adj} \rightarrow (\bm{3} \oplus \bm{7} \oplus \cdots \oplus \bm{2M-1}, \bm{1} \oplus \bm{S^2}) \oplus (\bm{1} \oplus \bm{5} \oplus \cdots \oplus \bm{2M-3}, \bm{A^2}) \,.
\end{equation}
Here, $\bm{S^2}$ and $\bm{A^2}$ are the 2-symmetric and 2-antisymmetric irreducible representations of $\mathfrak{f}_M = \mathfrak{so}(N/M)$. Similarly, when $M$ is odd the branching rule of the adjoint is
\begin{equation}
    \textbf{adj} \rightarrow (\bm{3} \oplus \bm{7} \oplus \cdots \oplus \bm{2M-3}, \bm{1} \oplus \bm{A^2}) \oplus (\bm{1} \oplus \bm{5} \oplus \cdots \oplus \bm{2M-1}, \bm{S^2}) \,,
\end{equation}
where $\bm{S^2}$ and $\bm{A^2}$ are now the 2-symmetric and 2-antisymmetric irreducible representations of $\mathfrak{f}_M = \mathfrak{usp}(N/M)$. Following the logic described in Section \ref{sec:nilphiggsing}, we now determine the contributions to the infrared anomalies from the Nambu--Goldstone modes. The Nambu--Goldstone anomaly coefficients are
\begin{equation}
  \begin{aligned}
    H &= \frac{1}{2} \bigg[ \frac{N^2(M-1)}{2M} \bigg] + \frac{N}{4} - \begin{cases} \frac{N}{4M} \quad &\text{if $M$ odd} \\
    0 \quad &\text{if $M$ even, } \end{cases} \\
    H_v &= \frac{1}{2} \bigg[\frac{N^2(M-1)(2M(M-1) - 1)}{6M}\bigg] - \frac{N(1 + 6M - 4M^2)}{12} + \begin{cases} \frac{N}{4M} \quad &\text{if $M$ odd} \\
    0 \quad &\text{if $M$ even, } \end{cases} \\
    H_M &= \begin{cases} N(M-1) + 2(M-1) \quad &\text{if $M$ odd} \\
    2N(M-1) + 4M \quad &\text{if $M$ even,} \end{cases}
  \end{aligned}
\end{equation}
where the terms in square brackets are, respectively, the $H$ and $H_v$ associated to $\mathfrak{su}(N)$ Higgsed by the partition $[M^{N/M}]$.

Putting together the UV anomalies with the contributions from the
Nambu--Goldstone modes, we work out the conformal
anomalies of the Higgsed SCFTs. The infrared central charges are
\begin{equation}\label{eqn:c,arecUSp}
    \begin{aligned}
	    12c^\text{IR} &= \frac{N^2p}{2} + \frac{Np}{2} + \frac{M^2N^2}{2p} - \frac{N^2}{2p} + \frac{M^2N}{p} - \frac{N}{p} \\&\qquad\qquad\qquad - MN^2 + \frac{N^2}{2M} - \frac{3MN}{2} + \begin{cases} \frac{N}{M}  &\quad\text{ if $M$ odd} \\ 0 &\quad\text{ if $M$ even, } \end{cases} \\
        48a^\text{IR} &= 2N^2p + 2Np + \frac{2M^2N^2}{p} - \frac{3N^2}{2p} + \frac{4M^2N}{p} - \frac{7N}{2p} \\&\qquad\qquad\qquad - 4MN^2 + \frac{3N^2}{2M} - 6MN + \begin{cases} \frac{7N}{2M}  &\quad \text{ if $M$ odd} \\ 0 &\quad\text{ if $M$ even. } \end{cases}
    \end{aligned}
\end{equation}
The difference of the central charges is
\begin{equation}\label{eqn:cmarecUSp}
        48(c^\text{IR} - a^\text{IR}) = \frac{N^2}{2M} - \frac{N}{2p} - \frac{N^2}{2p} + \begin{cases} \frac{N}{2M}  &\quad\text{ if $M$ odd} \\ 0 &\quad\text{ if $M$ even. } \end{cases}
\end{equation}
Finally, we turn to the enumeration of the infrared flavor algebras and the associated flavor central charges. When $M$ is even, the infrared flavor algebra is $\mathfrak{so}(N/M)$ with flavor central charge
\begin{equation}\label{eqn:kfrecUSpeven}
    k^\text{IR}_M = 2N - \frac{4M}{p} - \frac{2MN}{p} \,.
\end{equation}
Similarly, when $M$ is odd, the infrared theory has a flavor algebra which is
$\mathfrak{usp}(N/M)$ where the flavor central charge is
\begin{equation}\label{eqn:kfrecUSpodd}
    k^\text{IR}_M = N - \frac{2M}{p} - \frac{MN}{p} + 2 = (N+2)\frac{p-M}{p}\,.
\end{equation}

Now, we wish to compare these conformal anomalies to those obtained from the supergravity duals using the holographic dictionary. The line charge is the same as for $\mathfrak{so}(N)$ in equation \eqref{eqn:lineline}, except now we have
\begin{equation}
    b = N + 2 \,, \quad m = 2 \,, \qquad s_G = \frac{1}{2} \,,
\end{equation}
and the constant shift by $-2$ is replaced by a constant shift by $+2$. Specifically, the line charge is 
\begin{equation}
    \lambda(\eta)= 2 + \eta \left( \sum_{j=i}^N m_j - \frac{N+2}{p} \right) + \sum_{j=1}^{i-1} j m_j \,,
\end{equation}
for $\eta$ in the range $i - 1 \leq \eta \leq i$. The two-derivative contribution to $(2a - c)$ from the anomaly inflow is simply
\begin{equation}
  \begin{aligned}
    12(2a^\text{IR}-c^\text{IR})^\textrm{inflow} \big|_{2\partial} &= - \frac{3}{2} \int_0^p \lambda(\eta)^2 d\eta \\ 
    &= MN^2 - \frac{M^2 N^2}{2p}  - \frac{N^2 p}{2} - \frac{M^2 N}{p} + 3MN - 2Np  - 2p  \,.
  \end{aligned}
\end{equation}
We can see that this matches equation \eqref{eqn:c,arecUSp} at $\mathcal{O}(N^3)$. Turning to the difference of the central charges, we find
\begin{equation}
    48(c^\text{IR}-a^\text{IR})^\text{inflow} = -\frac{1}{2} \frac{N}{M} \left( \lambda(M) + \alpha_M \right) = - \frac{N^2}{2M} + \frac{N}{p} + \frac{N^2}{2 p} - \frac{N}{M}\left(1 + \frac{\alpha_M}{2}\right)\,.
\end{equation}
Recalling that $\alpha_M = -1$ if $M$ is odd, and $-2$ if M is even, we can see that this expression matches the field theory result in equation \eqref{eqn:cmarecUSp} at $\mathcal{O}(N^2)$ and $\mathcal{O}(N)$; furthermore, if we consider only the $M$-dependent terms, we match the expected result including $\mathcal{O}(1)$. Finally, we turn to the flavor central charges, which, to leading order, are given by
\begin{equation}
    k_M^\text{inflow} = -2 s_G^{(M)} \lambda(i) = \begin{cases} 
         -4 - 2 N + \dfrac{4M}{p} + \dfrac{2MN}{p} &\qquad \text{ if $M$ even,} \\[15pt]
         -2 - N + \dfrac{2M}{p} + \dfrac{MN}{p} &\qquad \text{ if $M$ odd.} \\
    \end{cases}
\end{equation}
In comparison to the field theory result, we find
\begin{equation}
    k_M^\text{IR} + k_M^\text{inflow} = \begin{cases}
        -4 &\qquad \text{if $M$ even,} \\
        0 &\qquad \text{if $M$ odd.} \\
    \end{cases}
\end{equation}
As we can see, this evidences agreement at $\mathcal{O}(N)$, and for all $M$-dependent terms; the mismatch only occurs at $\mathcal{O}(1)$, in such a way that does not depend on the choice of Higgsing.

\section{Discussion}\label{sec:disc}

In this paper, we constructed Lin--Lunin--Maldacena-type solutions of 11d supergravity that we propose are the holographic duals to 4d $\mathcal{N}=2$ Argyres--Douglas theories engineered via class $\mathcal{S}$ with irregular punctures. In particular, we incorporated various orbifold actions on the supergravity side, which correspond to starting with the 6d $(2,0)$ theory of $D$-type and/or incorporating twist lines in the construction of the dual field theory. To wit, we constructed the holographic duals to the theories
\begin{equation}\label{eqn:theparents}
    \mathcal{D}_{p}^{\,b}(SO(2N)) \,, \quad \mathcal{D}_{p}^{\,b}(SO(2N+1)) \,, \quad \mathcal{D}_{p}^{\,b}(USp(2N)) \,, \quad \mathcal{D}_{p}^{\,b}(USp'(2N)) \,.
\end{equation}

The $\mathcal{D}_p^{\,b}(G)$ theories possess a flavor symmetry $G$ and, for $p$ sufficiently large, the moment map of this flavor symmetry can be Higgsed by a choice of nilpotent orbit of $G$; in the class $\mathcal{S}$ language, this corresponds to the partial closure of the regular puncture. Part of the data of the supergravity solution is a choice of piecewise-linear function known as the line charge, $\lambda(\eta)$. The line charge is associated to an integer partition, and thus for each choice of nilpotent orbit for a classical Lie algebra there is a corresponding line charge. We propose that the Higgsed Argyres--Douglas theories obtained by nilpotent Higgsing of the $G$-flavor symmetry of the parent theories in equation \eqref{eqn:theparents} have holographic duals which are locally identical to those of the parent theories, except that there is global orbifolding and the line charge is modified according to the choice of nilpotent orbit by which the Higgsing is carried out. This provides a proposed holographic dual to the 4d $\mathcal{N}=2$ SCFTs known as
\begin{equation}
    \mathcal{D}_p^{\,b}(G, O) \,,
\end{equation}
for $G$ any classical simple Lie algebra and $O$ an arbitrary nilpotent orbit of $G$.

To validate this conjectured holographic duality, we matched various quantities that are calculable on both the supergravity and field theory side. First, we determine that the flavor symmetry algebra, as read off from the line charge on the supergravity side and from the nilpotent orbit on the field theory side, matches. Next, we determine the central charges, $a$ and $c$, as well as the flavor central charges for each simple non-Abelian factor in the flavor symmetry, from both perspectives, and we observe that they agree at leading or subleading order. While the line charge is naively associated to an integer partition, a nilpotent orbit for $G$ of $BCD$-type in fact corresponds only to integer partitions satisfying certain conditions. These conditions are automatically incorporated in the supergravity solution due to the $t$-rule, which must be satisfied for consistency when the solution involves orbifolding; that the $t$-rule precisely matches the requirement on integer partitions to correspond to nilpotent orbits of $BCD$-type algebras is another non-trivial check of the proposed duality. 

\paragraph{\uline{Further checks:}} In this paper, we carried out several checks of the proposed holographic duality, including matching the difference of the central charges at both leading and subleading orders. One further check is to determine $(2a - c)$ also to subleading order; here, we focused on the two-derivative contribution, however there are higher-derivative contributions coming from the $X_8$ term in M-theory. Incorporating these $X_8$ higher-derivative terms would allow us to test the duality at subleading order in $(2a-c)$. Another quantity which can be extracted from the supergravity dual is the conformal dimension of certain $1/2$-BPS operators. For the $\mathcal{D}_p(A)$ Argyres--Douglas theories studied in \cite{Bah:2021hei,Bah:2021mzw,Couzens:2022yjl, Bah:2022yjf}, the authors extracted the conformal dimensions of certain Coulomb branch operators, such as those appearing in equation \eqref{eqn:CBuntw}, by considering M2-branes wrapping cycles located at the kinks of the line charge. It would be interesting to reproduce the scaling dimensions, in equation \eqref{eqn:CBtw}, of the operators associated to the twisted Casimirs, which only occur for $\mathcal{D}_p(BCD)$. Moreover we have been unable to find a constraint which restricts $p$ to be an odd integer in the case of twisted punctures; reproducing this constraint from supergravity will provide a further check of the proposed duality. 

\paragraph{\uline{Holographic Higgs branch RG flows:}} For each choice of  $\mathcal{D}_p^{\,b}(G)$, where $G$ is a classical Lie algebra, we have constructed $\operatorname{AdS}_5$ solutions in 11d supergravity for each SCFT arising as the infrared fixed point after Higgsing the moment map of $G$ by an arbitrary nilpotent orbit of $G$. Field-theoretically, it is expected that if $O'$, $O$ are nilpotent orbits of $G$ and $O' < O$ under the partial ordering of nilpotent orbits, then there exists a Higgs branch renormalization group flow from the theory Higgsed by $O$ to the theory Higgsed by $O'$. It would be interesting to observe the existence of these RG flows from the holographic perspective. Much like in the 6d $(1,0)$ context studied in \cite{DeLuca:2018zbi}, we expect domain wall solutions of 11d supergravity that interpolate between the AdS$_5$ solutions with two different line charges, corresponding to nilpotent orbits $O$ and $O'$, to exist if and only if $O$ and $O'$ are ordered under the partial ordering of nilpotent orbits. This provides a significant further verification of our proposed holographic dictionary, as it would demonstrate that the structure of the interacting fixed points on the Higgs branch of the SCFT is replicated from the supergravity perspective. We plan to return to such an analysis in the future.

\paragraph{\uline{Holographic duals of \boldmath{$a = c$} SCFTs:}} Recently, an interesting class of 4d $\mathcal{N}=1$ and $\mathcal{N}=2$ SCFTs with identical central charges has been obtained via the diagonal gauging of a collection of $\mathcal{D}_p^{\,b}(G)$ theories \cite{Kang:2021lic,Kang:2021ccs,Kang:2022zsl,Kang:2022vab,Kang:2023dsa,LANDSCAPE}. Theories that have exactly $a=c$, without taking any form of large $N$ limit, are particularly interesting from the perspective of holography.  The difference, $(c-a)$, arises from the four-derivative correction to the effective action \cite{Anselmi:1998zb,Henningson:1998gx}, and there is no a priori reason for a such a contribution to vanish. In particular, there must be miraculous cancellations between the $(c-a)$ contributions of all the Kaluza--Klein modes arising from the compact sector of the supergravity solution. As these $a=c$ SCFTs were constructed via the diagonal gauging of $\mathcal{D}_p(G)$ theories, it is natural to attempt to mimic the gauging procedure on the holographic duals of $\mathcal{D}_p(G)$ that we have explored in this paper. The construction of the holographic duals to this class of $a=c$ theories, the determination of the Kaluza--Klein spectrum, and obtaining an understanding of the physical mechanism protecting $(c-a)$ is the subject of ongoing research.

\paragraph{\uline{Linear quivers from Type IIA:}} In this paper, we have focused on the holographic duals of the non-Lagrangian $\mathcal{D}_p(BCD)$ theories. Alternatively, there are 4d $\mathcal{N}=2$ Lagrangian SCFTs that live on D4-NS5-D6 intersecting brane configurations in Type IIA string theory. These generally take the form of linear quivers of $\mathfrak{su}$ gauge nodes. Such SCFTs are known to have holographic duals in Type IIA supergravity \cite{Gaiotto:2009gz,Reid-Edwards:2010vpm,Aharony:2012tz}; in fact these solutions are particularly interesting as they can be thought of as the local behavior, around a regular puncture, of the holographic solution of an arbitrary (untwisted type-A) class $\mathcal{S}$ theory. These solutions are circle reductions of LLM-type solutions, and also involve a line charge $\lambda(\eta)$, which is similarly related to a nilpotent orbit.  In the same way that we included orbifolding in this paper to understand the twisted punctures and the untwisted D-type punctures, we can incorporate orientifold four-branes in the holographic duals, as initiated in \cite{Nishinaka:2012vi}, to understand the local behavior around a twisted or D-type puncture. Such a process leads to orthosymplectic linear quivers, and we apply the careful analysis of the physical features via the holographic dictionary, developed in this paper, to extract properties of these quivers in \cite{CKLLquiver}. 

\paragraph{\uline{Spindles and $\mathcal{N}=1$ class $\mathcal{S}$:}}

In \cite{Ferrero:2020laf} the first spindle solution was given, and since then there has been much interest in constructing these solutions in different supergravity theories; for a sample of the recent literature, see, e.g.,  \cite{Ferrero:2020laf,Ferrero:2020twa,Hosseini:2021fge,Boido:2021szx,Faedo:2021kur,Ferrero:2021wvk,Cassani:2021dwa,Ferrero:2021ovq,Couzens:2021rlk,Faedo:2021nub,Ferrero:2021etw,Couzens:2021cpk,Giri:2021xta,Couzens:2022agr,Cheung:2022wpg,Suh:2022olh,Arav:2022lzo,Couzens:2022yiv,Couzens:2022aki,Boido:2022mbe,Suh:2022pkg,Inglese:2023wky,Suh:2023xse,Amariti:2023mpg,Kim:2023ncn,Hristov:2023rel}. A spindle is the weighted projective space $\mathbb{WCP}^1_{[n_{\pm}]}$, where $n_{\pm}$ are relatively prime, positive integers, and the supergravity solutions generically preserve minimal supersymmetry. Topologically a spindle is a two-sphere with conical deficits, with deficit angles $2\pi(1-n_{\pm}^{-1})$, at the two poles. The study of spindles developed in parallel to the study of disc solutions, however they are different global completions of the same local solutions \cite{Couzens:2021tnv, Couzens:2021rlk} and share many similar properties. One of the most striking properties is the necessity of mixing the R-symmetry with the isometry of the surface, in fact there are two types of twist \cite{Ferrero:2021etw}. Recall that for the class $\mathcal{S}$ Argyres--Douglas theories twisting the R-symmetry with the $\mathbb{CP}^1$ isometry is a necessary requirement of having an irregular puncture. Despite the large number of spindle supergravity solutions constructed, their field theory duals are not understood, and it would be interesting to remedy this in the future. As they share a common supergravity origin, it is tempting to speculate that there may be flows between the 4d $\mathcal{N}=1$ SCFTs dual to M5-branes on a spindle and the $\D_p(G)$ theories studied here.

\paragraph{\uline{Redundancy in the Argyres--Douglas landscape:}} A recent exploration of redundancy in the landscape of Argyres--Douglas theories has appeared in \cite{Beem:2023ofp}. In particular, the authors of \cite{Beem:2023ofp}, following on the work of \cite{Xie:2019yds}, proposed that
\begin{equation}\label{eqn:regHiggsed}
    \mathcal{D}_p\left(SU(N),  [1^{m_1}, 2^{m_2}, \cdots, N^{m_N}]\right) \,,
\end{equation}
and
\begin{equation}\label{eqn:overHiggsed}
    \mathcal{D}_p\left(SU(p\ell + N),  [1^{m_1}, 2^{m_2}, \cdots, N^{m_N}, p^\ell]\right) \,,
\end{equation}
are the same SCFT, for any positive integer $\ell$.\footnote{For simplicity, we assume here that $\gcd(p,N) = 1$ and $p > N$.} A priori, this is surprising as the latter theory appears to have an additional $SU(\ell)$ flavor symmetry, however, a straightforward application of equation \eqref{eqn:tachiGB} reveals that the infrared flavor central charge of the $SU(\ell)$ factor is
\begin{equation}
    k_p = p \left( \frac{2(p-1)}{p} (p\ell + N) \right) - 2\left((p-1)N + \ell p(p-1)\right) = 0 \,.
\end{equation}
Thus the $SU(\ell)$ does not act faithfully on the SCFT. In the holographic dual of this latter theory, we find a line charge of the form
\begin{equation}
    \lambda(\eta) = \eta \left( \sum_{j=i}^{N + p\ell} m_j - \frac{N+p\ell}{p} \right) + \sum_{j=1}^{i-1} j m_j \,.
\end{equation}
This is an identical line charge to that for the theory in equation \eqref{eqn:regHiggsed}, over the range of $\eta$ such that $\lambda(\eta) \geq 0$.  Thus, the local forms of the supergravity duals are identical. Some extra care needs to be taken with matching the global aspects of the solution, such as flux quantization, however, in the end, the redundancy of \cite{Beem:2023ofp} is apparent from the holographic duals. Given the understanding of this redundancy of $\mathcal{D}_p(A)$ theories from the line charge of the holographic dual, we expect that one can obtain putative redundancies in the Higgsed $\mathcal{D}_p(BCD)$ landscape by comparing the supergravity solutions that we have constructed here.

\section*{Acknowledgements}

We thank Pieter Bomans, Jacques Distler, Alessandro Mininno, Matteo Sacchi, and Palash Singh for useful discussions.  M.J.K., C.L., and Y.L.~thank the 2023 Simons Summer Workshop for providing a stimulating environment during an important stage of this work; M.J.K.~and C.L.~also thank Kyung Hee University for hospitality during an earlier stage of this project.
M.J.K.~is supported by the U.S.~Department of Energy, Office of Science, Office of High Energy Physics, under Award Numbers DE-SC0011632 and DE-SC0013528, and QuantISED Award DE-SC0020360, and a Sherman Fairchild Postdoctoral Fellowship.
C.L.~acknowledges support from DESY (Hamburg, Germany), a member of the Helmholtz Association HGF. Y.L.~is supported by the National Research Foundation of Korea under the grant NRF-2022R1A2B5B02002247.

\bibliographystyle{sortedbutpretty}
\bibliography{ref}
\end{document}